\documentclass[fleqn,usenatbib,linenumbers]{mnras}
\usepackage{newtxtext,newtxmath}

\usepackage[T1]{fontenc}
\usepackage{ae,aecompl}
\usepackage{graphicx}	
\usepackage{amsmath}	
\usepackage{amssymb}	

\title[]{A Two-Zone Model as origin of Hard TeV Spectrum in Extreme BL Lacs}
\author[E. Aguilar-Ruiz]{
E. Aguilar-Ruiz$^{1}$\thanks{E-mail: eaguilar@astro.unam.mx}, N. Fraija$^{1}$,  A. Galvan-Gamez$^{1}$ and E. Ben\'itez$^{1}$  \\
$^{1}$Universidad Nacional Aut\'onoma de M\'exico, Instituto de Astronom\'\i{}a, AP 70-264, CDMX  04510, Mexico \\}

\date{Started 25/11/2020. Finished YYYY; in original form ZZZ}

\pubyear{2022}

\begin{document}
\label{firstpage}
\pagerange{\pageref{firstpage}--\pageref{lastpage}}
\maketitle

\begin{abstract}
The emission of the so-called extreme BL Lacs poses challenges to the particle acceleration models. The hardness of their spectrum, $\lesssim 2$,  in the high-energy band demands unusual parameters using the standard one-zone synchrotron self-Compton (SSC) model with a deficient magnetized plasma. Some authors use either two-zone or hadronic/lepto-hadronic models to relax these atypical values. In this work, we present a lepto-hadronic two-zone model to explain the multi-wavelength observations of the six best-known \textit{extreme} BL Lacs. The very-high-energy gamma-ray observations are described by the photo-hadronic processes in a blob close to the AGN core and by SSC and external inverse Compton-processes in an outer blob. The photo-hadronic interactions occur when accelerated protons in the inner blob interact with annihilation line photons from a sub-relativistic pair plasma. The X-ray observations are described by synchrotron radiation from the outer blob. The parameter values found from the description of the spectral energy distribution for each object with our phenomenological model are similar to each other, and lie in the typical range reported in BL Lacs.
\end{abstract}

\begin{keywords}
astroparticle physics $-$  BL Lacertae objects: general $-$   BL Lacertae objects: individual: 1ES 0229+200 $-$ radiation mechanisms: non-thermal $-$ relativistic process $-$ $\gamma$-rays : galaxies. 
\end{keywords}

\section{Introduction}

Blazars are a class of active galactic nuclei (AGN) that present a relativistic jet pointing close to the line of sight \citep{Urry&Padovani1995}. Due to this particular orientation, blazar emission is dominated by relativistic aberration and boosting effects produced in the jet \citep{Blandofor_1979ApJ...232...34B}. They are multi-wavelength emitters from the radio to the gamma-ray bands and found to be the most numerous extragalactic gamma-ray sources  \citep{Ackerman_2015ApJ...810...14A}. The spectral energy distribution (SED) of blazars shows two low- and high-energy peaked humps. The low-energy hump peaks in the sub-mm to X-rays energy band, and the high energy at the MeV-TeV energies. Based on optical spectroscopy, blazars are usually divided into two classes: Flat Spectrum Radio Quasars (FSRQs) and BL Lac objects (BL Lacs) \citep[e.g.][]{Angel_1980ARA&A..18..321A, Padovani_2017A&ARv..25....2P}. The FSRQs present strong emission lines, whereas BL Lac objects have weak (i.e., equivalent width (EW) $<$ 5\,\AA) or absent emission lines \citep{Marcha_1996MNRAS.281..425M}. 
In particular, the BL Lac class is further classified depending on the location of the low-energy peak in the SED, which is well fitted with  synchrotron emission. Therefore, low-synchrotron peak (LSP) BL Lacs have peaks at $\nu^{S}_{p}\,<\,10^{14}\,{\rm Hz}$; intermediate-synchrotron peak (ISP) at  $10^{14}\,<\,\nu^{S}_{p}\,<\,10^{15}\,{\rm Hz}$, and high-synchrotron peak (HSP) at $\nu^{S}_{p}\,>\,10^{15}\,{\rm Hz}$ \citep{Abdo_2010ApJ...716...30A}. 

The existence of diversity in the positions of the synchrotron peak and the jet power hints at a possible correlation between them. \citet{1998MNRAS.299..433F} proposed the existence of the so-called blazar sequence where the LSP BL Lacs and FRSQ are found to be the most powerful sources, while the HSP BL Lacs appeared as the less powerful ones. Also, a new type of HSP BL Lacs has an intrinsic very-high-energy (VHE) spectrum ($\Gamma\,\leq\,$1.5-1.9) with their peak located above $>2-10\rm\,TeV$. These are the so-called \textit{extreme}-TeV BL Lacs \citep[TBLs;][]{Tavecchio&etal_2011, Costamante2018}. Regarding the synchrotron peak, most have a high peak above 1 keV ($>10^{17} \, {\rm Hz}$). Due to the extreme or high position of their synchrotron peak, \cite{2001A&A...371..512C} named them as \textit{Extreme}-HSP (EHSP). 

A blazar can be an EHSP and a TBL simultaneously, but there are no pieces of evidence that support the necessary co-existence of the two conditions. On the other hand,  during its lifetime, a blazar can show only EHSP or TBL behavior, suggesting the extreme condition is a transient stage, e.g., \citep{Ahnen2018}. Furthermore, \cite{Foffano&etal_2019} proposed that the EHSP objects are not a homogeneous class because they show different spectral properties in the TeV-band.

Modeling the low-energy peak of EHSP objects with synchrotron emission from an electron population does not represent a problem by the standard shock acceleration mechanisms considering one-zone geometry \citep[e.g.][]{Ahanorian_2004vhec.book.....A, Tavecchio_2010MNRAS.401.1570T, 2017APh....89...14F, 2017ApJS..232....7F}. However, when using the standard one-zone synchrotron self-Compton (SSC) model, there appear many problems when modeling the high-energy TeV peak \citep[see][for a recent review]{Biteau_2020NatAs...4..124B}. In the case of TBLs, these models demand parameters with exceptional values \citep[e.g., see][and references therein]{Costamante2018, 2015MNRAS.448..910C}. One of the main problems is that TBL sources operate in the Klein-Nishina (KN) regime, for which the flux is expected to be suppressed, thus making the formation of a hard-spectrum difficult \citep{2009MNRAS.399L..59T}.  Another issue is that the SSC model demands a weakly magnetized plasma with an equipartition value much less than the unity ($U_B/U_e \ll 1$), which contradicts the observations \citep{Nemmen&etal_2012}.  Moreover, SSC model in TBL requires a high Doppler boosting factor (e.g., $\mathcal{D} \geq 40$) and minimum electron Lorentz factors as great as $\gamma_{e,\rm min} \gtrsim 10^3$ \citep[e.g.][]{Kaufmann2011, 2009MNRAS.399L..59T, 2006MNRAS.368L..52K}.

Different models have been proposed to relax the parameter values required by the SSC model. For example, \cite{Aharonian&etal_2008} invoked $\gamma\gamma$-absorption with a narrow Planck radiation field covering the emission zone where VHE gamma-rays are produced. \cite{Lefa2011} showed that a harder spectrum could result if adiabatic losses dominate synchrotron losses or if a Maxwellian-like electron distribution is considered.  Another proposal recently explored by \cite{Tavecchio&Sobacchi_2020} mentioned that energetic electrons could not be fully isotropised, resulting in an anisotropic distribution and a harder SSC spectrum with an equipartition value close to unity.  On the other hand, either hadronic (proton synchrotron) or lepto-hadronic (induced cascade) models also have been proposed to relax the extreme parameter's values demanded by the SSC model, \citep[e.g.,][]{2015MNRAS.448..910C, Tavecchio_2014, Zheng&etal_2016}.

In this work, we propose a lepto-hadronic two-zone model to explain the observed broadband spectrum for \textit{extreme} BL Lacs (i.e., those with EHSP and TBL simultaneous behavior) to avoid the extreme conditions demanded by SSC one-zone model. Our model takes two zones described by the standard blob geometry \citep[e.g.,][]{Gloud_1979A&A....76..306G,Ghisellini_1989MNRAS.236..341G, Ghisellini_1998MNRAS.301..451G, Tavecchio_1998ApJ...509..608T, Katarzynski_2001A&A...367..809K}. The first zone is responsible for the highest gamma-rays peaking at TeV energies produced by the photo-hadronic processes, and the second zone accounts for the X-rays and sub-TeV gamma-rays. In our model, external photon sources are considered, specifically from the broad-line region (BLR) and the dusty torus (DT), which significantly contribute to the observed SED due to their interactions with accelerated electrons in the outer blob via external Compton (EC) models  \citep[e.g.][]{Dermer_1992A&A...256L..27D, Dermer_1993ApJ...416..458D, Sikora_1994ApJ...421..153S, Blazejowski_2000ApJ...545..107B, Donea_2003APh....18..377D, Finke_2016ApJ...830...94F, 2019ApJS..245...18F}. In some previous studies, it has been found that the emission from BLR/DT is lacking in BL Lac objects \citep[e.g.][]{2012ApJ...745L..27P}. However, some HSP BL Lacs like Mrk 421, PKS 2005-489, PKS 2155-304, and H 2356–309 present weak Ly$\alpha$ emission with an emission's line luminosity of the order of Ly$\alpha \sim 10^{41}-10^{42} \, \rm erg \, s^{-1}$ \citep{Sbarrato_2012MNRAS.421.1764S,Stocke_2011ApJ...732..113S, Fang_2014ApJ...795...57F}.   Moreover, considering the result introduced by \cite{Elitzur_2009ApJ...701L..91E} we note that the emission from BLR/DT only disappear if their disk luminosity is lower than the critical value $L_{d, \rm crit} \approx 1.1 \times 10^{41} \, {\rm erg \, s^{-1}} (2.3 \times 10^{40} \, {\rm erg \, s^{-1}})$ for $M_\bullet = 10^9 (10^8) M_\odot$. 
Then as \cite{Foschini_2012RAA....12..359F} pointed out, these BL Lacs have standard disks similar to those operating in FSRQs but with less ionization efficiency. Therefore, although EC models are not usually invoked to fit the SED in BL Lacs, especially in the HSP types, they provide a feasible explanation for some BL Lacs where a pure SSC model requires extreme parameters to explain its broadband emission.  For example, \cite{Kang_2016MNRAS.461.1862K} used a one-zone leptonic model with a SSC and EC scenario for the particular case of the HSP BL Lac PKS 1424 +240.  Using both a SSC and EC model, these authors found a set of parameters that successfully fit the observations.  They suggested that the energy densities of the BLR/DT must be lower than the required by traditional FSRQs about a factor of $\sim 10^{-3}$.   Also, the authors suggest that the BLR/DT do not disappear entirely in BL Lacs but are present with weaker intensity. Furthermore, \cite{Cerruti_2017A&A...606A..68C} pointed out that in the case of leptonic scenarios, the two-zone and the EC model are favored to explain the electromagnetic emission of PKS 1424 +240. The DT radiation is the most significant external photons source for the latter.

Moreover, we consider photons from an annihilation line produced by an outflowing electron-positron plasma that emerges above the accretion disc \citep{1999MNRAS.305..181B} (hereafter named as pair-plasma). This plasma will be the most important radiation source to interact with the accelerated protons in the inner blob, and to form a hard spectrum in gamma-rays which extends up to the TeV regime.  

This work is structured as follows: In section 2, we describe in detail our proposed model. In section 3, we apply our model to six extreme BL Lacs. Also, analytic estimations are presented to constrain the parameters scan and to model the SED. Moreover, the resulting SED for the six selected extreme BL Lacs is shown. Finally, in section 4, we present a discussion of our results and their implications.  In this work our cosmology is based on a $\Lambda$CDM model where $H_0 = 67.4 \; \rm km \, s^{-1} \, Mpc^{-1}$, and with energy densities corresponding to matter and dark energy $\Omega_M = 0.315$ and $\Omega_\Lambda = 0.685$, respectively \citep{2020A&A...641A...6P}.

\section{Model description}\label{EHBL:Multizone}
\begin{figure}
		{
			\begin{minipage}{\textwidth}
	    	\resizebox*{0.45\textwidth}{0.24\textheight}
				{\includegraphics{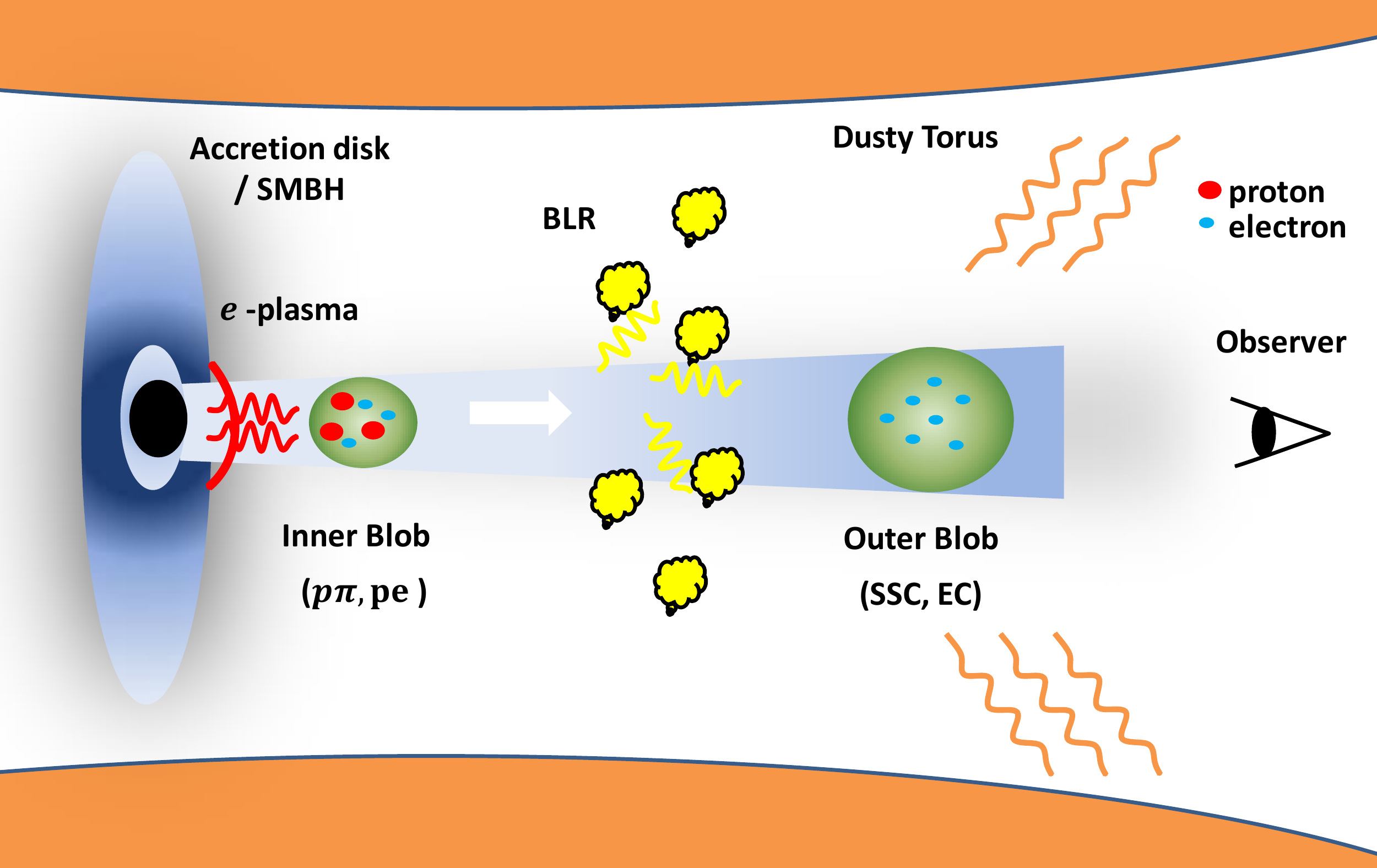}}
			\end{minipage}
		}
		\caption{A general picture (not at scale) of our proposed model is shown. We consider two relativistic blobs: an inner blob close to the SMBH and a giant outer blob far away from the central engine and beyond the BLR. Besides the two blobs, we consider three zones as sources of seed photons: i) gamma-ray  photons (the annihilation line-shape spectrum at 511 keV) from a pair plasma, ii) ultraviolet (UV) photons from the BLR, and finally, iii) the infrared (IR) radiation from the DT.}
		\label{Fig_model_skect}
\end{figure}

We propose that the hard gamma-ray spectrum can be described by the contribution of two different emission zones. We consider that relativistic electrons and protons are injected  in the first region called the ``inner blob", which is located at a distance $r_{\rm i}$ from the supper-massive black hole (SMBH). Accelerated protons interact with the radiation coming from the pair-plasma\footnote{The pair plasma is produced above the accretion disc by the annihilation of MeV photons and moves vertically.} via photopair ($pe$) and photopion  ($p\pi$) processes. It is worth noting that  although electrons must co-exist with protons in this region, we only consider hadronic interactions for modeling the source emission.  The pair-plasma created close to the SMBH at a very compact region releases photons with a characteristic energy at $\sim \rm 511 \, keV$. At a distance, $r_{\rm o}$ from SMBH, the second region of interest called the ``outer blob" is located beyond the BLR but immersed in the DT radiation field. We only consider leptonic (electrons) processes via SSC and EC for modeling the source in this region.  A general picture of our proposing model is sketched in Figure \ref{Fig_model_skect}.

In our model, the inner and the outer blobs can be described as a standard blob, which assumes a spherical geometry with size $R_{\rm b}$ moving relativistically with a Lorentz factor $\Gamma_i$ and $\Gamma_o$, respectively. The pair-plasma has a mildly-relativistic speed, i.e., with $\Gamma\sim$2, and the BLR and the DT regions are at rest. 

This work defines four reference frames: the pair-plasma, the inner and outer blobs, and the observer's frames. Here, we denote with Latin capital letter and the superscript ``${\rm ob }$" for observed quantities. At the same time, the AGN frame will also be denoted by the Latin capital letter but without the superscript. The plasma, the inner, and the outer blobs are marked with Greek lowercase letters with unprime, prime, and two-prime, respectively.  For example, the observed energy is written as $E^{\rm ob}$, and the energy measured in the comoving frame of the AGN, the pair-plasma, the inner blob, and the outer blob are $E$, $\varepsilon$, $\varepsilon^\prime$, $\varepsilon^{\prime\prime}$, respectively.

For the relativistic blobs, we consider the on-axis case when the viewing angle and  the Doppler factor are $\theta_{\rm obs} \lesssim 1/\Gamma$ and $\mathcal{D} = \left[ \Gamma(1-\beta\cos{\theta_{\rm obs}})\right]^{-1}$, respectively. We assume different speeds for the pair-plasma ($\Gamma_{\rm pl}$), and the inner ($\Gamma_{\rm i}$) and the outer ($\Gamma_{\rm o}$) blobs.  Therefore, each blob moves relative to the other one with a Lorentz factor
\begin{equation}\label{eq_GammaRel}
\Gamma_{\rm rel} = \Gamma_{\rm j} \Gamma_{\rm k} \left( 1 - \beta_{\rm j} \beta_{\rm k} \right)\,\,\,\,{\rm with}\,\,\,\, {\rm j\neq k}\,.
\end{equation}
The total power of the jet is the sum of each blob, and each blob carries energy in the form of radiation, particles, and a magnetic field. The total luminosity of each blob in the AGN frame is \cite[e.g.;][]{Celotti&Ghisellini_2008, Bottcher&etal_2013}
\begin{equation}
    L_{\rm j} = \pi R_{\rm j}^2 \, \Gamma_{\rm j}^4 \beta_{\rm j} c \, (u_{\rm \gamma, j} + u_{\rm e, j} + u_{\rm p, j} + u_{\rm B, j}) \, , 
\end{equation}
where $u_{\rm \gamma, j}$ is the radiation energy density given by $u_{\rm \gamma, j} \simeq L_{\rm \gamma, j}^{\rm ob}/(4\pi r_{\rm j}^2 \mathcal{D}_{\rm j}^2 \, c )$, with $L_{\rm \gamma, j}^{ob}$ the bolometric observed radiation.  
The terms $u_{\rm e,j}$ and $u_{\rm p,j}$ correspond to the particle energy densities for electrons and protons, respectively, defined as $u_{\rm i, j} = m_{i} c^2 \int \gamma_{\rm i, j} N_{\rm i, j}(\gamma_{i, j}) d\gamma_{\rm i, j}$ with ${\rm i} = e$ and $p$. Finally, the magnetic energy density is $u_{\rm B, j} = {B_{\rm j}^2}/(8\pi)$. In the following, we describe the external radiation sources relevant for the interaction with the inner and outer blob.

\subsection{External radiation sources}

\subsubsection{Annihilation line}

Near the SMBH, MeV-photons annihilate with lower energy photons ( i.e.,  $\gamma+\gamma \rightarrow e^\pm$), producing an $e^\pm$ outflow that moves with velocity $\beta_{\rm pl} \approx 0.3 - 0.7$. These pairs produce an optically thick environment if the luminosity above $\rm 511 \, keV$  is greater than $L_{\rm keV} \gtrsim 3 \times 10^{-3} \, L_{\rm Edd}$, where $L_{\rm Edd} \approx 1.26 \times 10^{47} \, {\rm erg\,s^{-1}}  \, ( M_{\bullet}/10^9 M_{\odot} )$ is the Eddington luminosity and $M_{\bullet}$ is the mass of the SMBH. Radiation only escapes at the photosphere's radius, $R_{\rm ph}$,  producing a line shape peaking at $\varepsilon_{\rm pl} \approx 511 \, \rm keV$ \citep{1999MNRAS.305..181B}. The photosphere occurs at a compact radius of the order of Schwarzschild radius $R_{\rm g} \approx 1.5 \times 10^{14} \, {\rm cm} \, ( M_{\bullet}/10^9 M_{\odot})$, and the released radiation is very anisotropic where the greater contribution is confined within a solid angle $\Omega_{\rm pl} \lesssim 0.2 \pi$. Here, we assume the emission's shape is very narrow such that a delta approximation can describe its photon distribution as \citep{Fraija&etal_2020MNRAS.497.5318F}

\begin{equation}\label{eq_line_distribution}
    n_{\rm pl} (\varepsilon) = \frac{u_{\rm pl} }{\varepsilon_{\rm pl}}\, \delta(\varepsilon - \varepsilon_{\rm pl})\,,
\end{equation}

where the normalization constant is $u_{\rm pl} = L_{\rm keV} / (\Omega_{\rm pl} R_{\rm ph}^2 \beta_{\rm pl } c)$ with $c$ de speed of light.

\subsubsection{The broad-line region} 
Photons coming from the BLR have contributions mainly from two emission lines corresponding to H I $\rm Ly \alpha$ and He I $\rm Ly \alpha$ with energies peaking at $E_{\rm BLR} = 10.2\, {\rm eV}$ and $ 40.2\, {\rm eV}$, respectively. This emission, spatially distributed in a thin shell geometry, is thought to the result of the scattering of UV photons from the accretion disk into clouds located at a distance of  \citep[e.g.][]{Ghisellini_2009MNRAS.397..985G, Kaspi_2007ApJ...659..997K}
\begin{equation}\label{eq_rBL}
    r_{\rm BLR} \approx 10^{17} \, {\rm cm} \; \left( \frac{L_{\rm d}}{10^{45} \, \rm erg \, s^{-1}} \right)^{1/2} \,.
\end{equation}
For BL Lacs, the bolometric BLR's luminosity has an upper limit given by the relation $L_{\rm BLR} \lesssim 5 \times 10^{-4} L_{\rm Edd}$  \citep{Ghisellini&etal_2011}, and if this is a fraction of the disc luminosity $L_{\rm BLR} = \phi_{\rm BL} L_d$, then the BLR energy density becomes \citep{Hayashida&etal_2012} 
\begin{equation}\label{eq_DTenergy}
    u_{\rm BLR} = \frac{\phi_{\rm BLR} L_{d}}{4\pi r_{\rm BLR}^2 c \left[ 1 + (r_o/r_{\rm BLR})^3 \right]} \approx 0.26  \, {\rm erg\, cm^{-3}} \; \frac { \phi_{\rm BLR}} { 1 + (r_o/r_{\rm BLR})^3 }\, .
\end{equation}
The photon distribution of the emission's line in the delta-approximation is given  by
\begin{equation}\label{eq_nBL}
    n_{\rm BLR} (E) = \frac{u_{\rm BLR}}{E_{\rm BLR}} \, \delta(E - E_{\rm BLR}) \, .
\end{equation}

\subsubsection{Dusty torus} 
Photons coming from the DT are thought of as reprocessing radiation from the accretion disc. Therefore, we can calculate its luminosity as $L_{\rm DT} = \phi_{\rm DT} L_d$ \citep[e.g.,][]{Ghisellini&Tavecchio_2008}.  The radius of this region can be calculated using the result of \cite{Cleary&etal_2007}, which is given by
\begin{equation}\label{eq_rDT}
    r_{\rm DT} \approx 2.5 \times 10^{18} \, {\rm cm} \; \left( \frac{L_{\rm d}}{10^{45} \, \rm erg \, s^{-1}} \right)^{1/2} \, ,
\end{equation}
where the emission can be modeled as a grey-body with a temperature of 100-1000 Kelvin and a frequency peak at $ \nu_{\rm DT} = 2\times10^{13} \, {\rm Hz}$. If the above relation holds for any blazar, then we observe that the photon energy density is independent of the disc luminosity \citep{Hayashida&etal_2012}
\begin{equation}\label{eq_DTenergy}
    u_{\rm DT} = \frac{\phi_{\rm DT} L_{d}}{4\pi r_{\rm DT}^2 c \left[ 1 + (r_o/r_{\rm DT})^3 \right]} \approx 4.3 \times 10^{-3}  \, {\rm \frac{erg}{cm^{3}}} \; \frac { \phi_{\rm DT}} { 1 + (r_o/r_{\rm DT})^4 }\, ,
\end{equation}
where the photon distribution is normalized by $ u_{\rm DT} =\int dE \, E \, n_{\rm DT}(E) $.

\subsection{The outer blob}
We assume that this blob is far away from the pair-plasma compared with the inner blob; then, IC scattering is not expected to be relevant due to a dilution of the photon number density. However, the radiation from the DT plays an important role because it interacts with the electrons in the blob. Because the outer blob is assumed to be located slightly out from the BLR, but inside the DT radiation field, the photons from the BLR are seen in the outer blob with redshifted energy of $\varepsilon_{\rm BLR}'' = E_{\rm BLR} /  2 \Gamma_{o} $ and the energy density is $u_{\rm BLR}'' = u_{\rm BLR} / 4 \Gamma_{o}^2$, whereas the photon energy and the energy density from the DT are blueshifted $\varepsilon_{\rm DT}'' = 2 \Gamma_{o}  E_{\rm DT} $ and $u_{\rm DT}'' = 4 \Gamma_{o}^2 u_{\rm DT}$. 
Furthermore, electrons inside this blob are cooling by the SSC mechanism producing the whole X-ray component and a fraction of the high-energy gamma rays. The electron distribution in this region follows the same spectral shape as the inner blob (see equation \ref{eq_eDistritbution}), but with an electron energy break reaching energies close to TeV energies ($\gamma_e \sim 10^{6} $). We neglect the proton contribution from this region because BLR and DT radiations only produce gamma-rays from photopion decay in the PeV regime with a lower flux.

In this blob, electrons are assumed with an homogeneous and isotropic distribution given by a broken power-law function 
\begin{equation}\label{eq_eDistritbution}
N_e'(\gamma_e') = K_e'
\begin{cases}
{\gamma'_e}^{-\alpha_{e,1}}, 
\qquad \qquad \qquad \quad \;\; {\gamma'_{e, \rm min}} \leq {\gamma'_e} \leq {\gamma'_{e, \rm br}} 
\\
{\gamma'_{e, \rm br}}^{\alpha_{e,2}-\alpha_{e,1}} {\gamma'_e}^{-\alpha_{e,2}}, 
\qquad {\gamma'_{e, \rm br}} \leq {\gamma'_e} \leq {\gamma'_{e, \rm max}}
\end{cases}
\end{equation}
where ${\gamma'}_{e, \rm min}$, ${\gamma'}_{e, \rm br}$ , ${\gamma'}_{e, \rm max}$ are the minimum, break and maximum Lorentz factor of ultrarelativisc electrons, respectively. These electrons immerse in a magnetic field emit radiation by synchroton mechanism. Electrons also interact with the radiation field coming from the pair-plasma by IC.

\subsubsection{Synchrotron radiation}
In a magnetic field, charged particles move along the field lines losing energy via synchrotron radiation. The power radiated by an electron distribution is \citep{BLUMENTHAL&GLOUD_1970, Finke_2008, Sauge_2004}
\begin{equation}\label{eq_syn_emission}
    J_{\rm s}'' (\varepsilon'') =  \frac{\sqrt{3} e^3 B''}{ 2 \pi \hbar m_e c^2}  \int_{ {\gamma''}_{e,{\rm min}} }^ { {\gamma''}_{e,{\rm max}} } d\gamma_e'' N''_e (\gamma_e'') \, R_{s}(x) \, ,
\end{equation}
where $e$ is the electron charge, $\hbar$ is the Plank's constant,  $x = \varepsilon''/\varepsilon''_c$, the characteristic energy is given by $\varepsilon''_c= \frac{3 e B'' \hbar }{2 m_e c} {\gamma''_e}^ 2 \sin{\theta}$ and ${\theta}$ is the pitch angle, the function $R_{s}$ is given in \cite{Finke_2008}.
\subsubsection{Inverse-Compton scattering}
An electron moving inside a radiation field produces Compton scattering. As well, synchrotron electron energy loss for each collision produces high-energy photons. The total emissivity coefficient produced by an isotropic electron population is \citep {BLUMENTHAL&GLOUD_1970}
\begin{equation}
J''_{ic} (\varepsilon''_c) = \frac{3}{4} c \sigma_{\rm T} \varepsilon''_c \int_{\frac{\varepsilon''_c}{m_e c^2}} d\gamma''_e \frac{N''_e(\gamma''_e)}{{\gamma''_e}^2}  \int d\varepsilon'' \frac{n''_{\rm ph}(\varepsilon'')}{\varepsilon''} F_c(q,\Gamma_e) \, .
\end{equation}

Finally, the IC radiation is calculated using the above equation but with a different photon spectrum for SSC and EC emission. The corresponding synchrotron emission for SSC is $n''_s(\varepsilon''_s) \approx J''_{s}(\varepsilon'')/ (4\pi {R_o''}^{2} c \varepsilon''_s)$ and for EC is the DT radiation with a greybody type distribution.

\subsection{The inner blob}

This blob is located very close to the pair-plasma photosphere; therefore, its size must be a few times $R_{\rm g}$.
We assume that the inner blob is electrically neutral \citep[i.e. $N_e=N_p$;][]{Bottcher_2013ApJ...768...54B, Sikora_2009ApJ...704...38S, Abdo_2011ApJ...736..131A, Petropoulou_2014A&A...562A..12P, Fraija_2017APh....89...14F}, and therefore, an energy density between protons and electrons as  $U_e \sim \frac{m_e \log\gamma_{e,\rm br}}{ m_p \log\gamma_{p,\rm max}} \, U_p$, for a spectral index $\approx 2$.

This indicate that the relativistic hadronic energy dominates over the leptonic one. Here, we assume protons are distributed homogeneously and isotropically  during the equilibrium stage, given by 
\begin{equation}\label{eq_pDistribution}
{N_p'}({\varepsilon_p'}) = {K_p'} {\varepsilon'_p}^{-\alpha_{p}} \, 
\qquad \varepsilon'_{p, \rm min} \leq \varepsilon_p' \leq \varepsilon'_{p, \rm max}\,,
\end{equation}
where $K_p'$ is the normalization constant, $\alpha_p$ is the proton spectral index and ${\varepsilon}'_{p, \rm min}$ and ${\varepsilon}'_{p, \rm max}$ correspond to the minimum and maximum energies in the comoving frame, respectively.
Due to relativistic effect photons between the pair-plasma and the inner blob frames (electromagnetic emission from the pair-plasma are redshifted in the inner blob frame), the energy and energy density become \citep[see,][]{Fraija&etal_2020MNRAS.497.5318F}

\begin{equation}\label{eq_plasma_energy}
   \varepsilon^{\prime}_{\rm pl} \approx {\varepsilon_{\rm pl}}/(2\Gamma_{\rm rel}) \, , \qquad {\rm and}\qquad   u^{\prime}_{\rm pl} \approx u_{\rm pl} /(2\Gamma_{\rm rel})^2\,,
\end{equation}
respectively, where $\Gamma_{\rm rel}$ is the relative Lorentz factor between the pair-plasma and the inner blob (see equation \ref{eq_GammaRel}). Therefore, we expect the interaction between accelerated protons with photons coming from the pair-plasma via photohadronic processes, i.e., photopion and photopair interaction, to produce gamma-rays peaking at TeV energies besides secondary leptons as electrons and neutrinos \citep{Fraija&etal_2020MNRAS.497.5318F}.

\subsubsection{Photopion production}
The presence of a radiation field produces interactions with accelerated protons via photopion process. The main contribution comes from $\Delta$-resonance, $p+\gamma \rightarrow p + \pi ^0 (n + \pi^+)$. The final stable particles resulting from pion decay are $\pi^0 \rightarrow 2\gamma$  and $\pi^+ \rightarrow \nu_\mu + \Bar{\nu}_\mu  + \nu_e + e^+$. From relativistic kinematics, the energy threshold to pion production in the head-on collision with pair-plasma's photons is
\begin{equation}
{\varepsilon'}_{p,\rm th}^{p\pi} = \frac{ m_\pi^2 + 2 m_\pi m_p }{4 \varepsilon'} \approx 70 \, {\rm PeV} \, \left( \frac{\varepsilon'}{\rm eV} \right)^{-1} \, ,
\end{equation}
where $m_\pi$ is the pion mass, $m_p$ is proton mass and $\varepsilon'$ is the photon target in the comoving frame.

The production rate of stable particles is given by \citep{Kelner&Aharonian_2008}
\begin{align}
{Q'}^{p\pi}_{i}(\varepsilon'_{i}) 
&= \int \frac{d\varepsilon'_p}{\varepsilon'_p} N'_p(\varepsilon'_p) \,\int d\varepsilon' \, n'_{\rm pl}(\varepsilon') \, \Phi_{i} (\eta,x),
\end{align}
with $\eta = \frac{4\varepsilon'_p\varepsilon'}{m_p^2 c^4}$ and $x=\frac{\varepsilon'_i}{\varepsilon'_p}$. The function $\Phi$ contains the physic of the process and depends on the type of particle to be considered. The label ${\rm i}$ represents $\gamma,\nu, \Bar{\nu},e^-,e^+$. 
The energy loss timescale is calculated considering all the particle type interactions as \citep{Kelner&Aharonian_2008}
\begin{align}
    t_{p\pi}^{-1} (\varepsilon'_p)&= \int_0^1dx \, x \, \int_{\varepsilon'_{min}}^{\infty}  d\varepsilon' \, n_{pl}(\varepsilon') \sum_{\rm all} \Phi_{i} (\eta,x) \, ,
\end{align}
with $\varepsilon'_{min}=\frac{\eta_0 m_p^2 c^4}{4 \varepsilon'_p}$ and $\eta_0 \approx 0.303$.
\subsubsection{Photopair production}
Another mechanism that the proton interacts with the radiation field is via photopair production, $p + \gamma \rightarrow p + e^- + e^+$ , i.e, Bethe-Heitler process. From relativistic kinematics, The energy threshold to pion production in the head-on collision is
\begin{equation}
{\varepsilon'}_{p,\rm th}^{pe} = \frac{ m_e^2 + m_e m_p }{\varepsilon'} \approx 479 \, {\rm TeV} \, \left( \frac{\varepsilon'}{\rm eV} \right)^{-1},
\end{equation}
and the electron production rate for photopair process in the case of $\delta$-approximation, assuming an isotropic photon distribution is given by  \citep{Romero&Vila_2008}
\begin{align}
Q'_e(\gamma'_e) \simeq \frac{6 }{16 \pi }\sigma_{\rm T} \alpha_f m_p m_e^2 \, \frac{N'_p(\gamma'_e m_p)}{{\gamma'_e}^{2}}   \; \int_{2m_e/\gamma'_e}^\infty d\varepsilon' \,  \frac{n'_{\rm pl}(\varepsilon')}{{\varepsilon'}^2}\, \psi\left(\kappa\right),
\end{align}
where the mean energy's fraction carried by an electron from the parent proton is approximated as $x \approx m_e/m_p$ (i.e, $\gamma'_e \approx \gamma'_p$) , $\alpha_f$ is the fine structure constant,  $\kappa=\frac{2\gamma'_p \varepsilon'}{m_e}$.
Finally, the energy loss timescale is
\begin{align}\label{eq_pe_loss}
t_{pe}^{-1}(\gamma'_p) = 
\frac{3 \sigma_{\rm T} \alpha_f m_e^3 }{16 \pi m_p }  \frac{1}{{\gamma'_p}^2} \int_{2m_e/\gamma'_p}^\infty d\varepsilon' \,  \frac{n'_{\rm pl}(\varepsilon')}{{\varepsilon'}^2} \varphi \left( \frac{2 \gamma'_p \varepsilon'}{m_e} \right)\,.
\end{align}
The functions $\psi$ and $\varphi$ are parametrized following  \cite{Chodorowski_1992}.
\subsubsection{Proton synchrotron}
Additionally to photohadronic processes, protons radiate via synchrotron if the magnetic field strength is enough to have a significant flux contribution. The flux from proton synchrotron is calculated using the equation (\ref{eq_syn_emission}) but replacing $m_e$ by $m_p$ \citep{Dermer&Menon2009}.
\subsubsection{Gamma-ray spectrum emerging from the inner blob}
Annihilation-line photons will attenuate the gamma-ray spectrum produced in the inner region from the pair-plasma and the radiation produced in the outer blob, BLR, and DT when these are passing through them. The optical depth for $\gamma\gamma$ attenuation is calculated following \citep{Stecker1992ApJ...390L..49S, Dermer&Menon2009}

\begin{equation}
    \tau_{\gamma\gamma}(E_\gamma) \simeq  \frac{l}{2c}\int_{-1}^{+1} d\mu (1-\mu) \int_{\varepsilon_{th}}^\infty d\varepsilon \; \sigma_{\gamma \gamma}(\beta_{\rm cm}) \; n_{\rm ph}(\varepsilon) \, ,
\end{equation}
where $\sigma_{\gamma\gamma}$ is the cross-section, $\mu$ the collision angle, and $l$ is the mean distance traveled by photons. Depending on the seed-photon source, the term $l$ corresponds to the size of the inner blob, the width of the shell of the BLR, and the size of the DT.
The threshold energy for pair creations and the created pair's velocity in the center of the mass frame is given by
\begin{equation}
    \varepsilon_{\rm th}=\frac{2 m_e^2 c^4}{E_\gamma (1-\mu)} \, ,\qquad {\rm and} \qquad \beta_{\rm cm} = \sqrt{1 - \frac{2 m_e^2 c^4}{E_\gamma\varepsilon (1-\mu)}} \, ,
\end{equation}
respectively.

\section{Application: Hard TeV BL Lacs}\label{EHBL:Application}

\subsection{Parameter selection and analytic estimations}
We consider the observational quantities, estimate the parameter values and apply the lepto-hadronic model with the two emission zones described in Section \ref{EHBL:Multizone} to model the broadband SED of the six known \textit{extreme} BL Lacs.  The observational quantities used for each extreme BL Lac are listed in the Table \ref{tab_EHBL_list}.

\subsubsection{The pair-plasma}

We set the photosphere's velocity as $\beta_{\rm pl, ph} \approx 0.3$ \citep{1999MNRAS.305..181B} and the emission line peaking at $\varepsilon_{\rm pl} \approx 511 \, {\rm keV}$ with a photon distribution given by equation  (\ref{eq_line_distribution}). We expect the formation of an outflowing pair-plasma when the luminosity is $L_{\rm keV} \gtrsim 3 \times 10^{-3} L_{\rm Edd}$, as previously mentioned. In addition, as the pair-plasma will be produced by the radiation of the accretion disc, we assume $L_{\rm keV} \sim L_d$.  However, if the ratio of BLR and disc luminosities, as generally accepted, is $L_d = 10 \, L_{\rm BLR}$, then the disc luminosity cannot be greater than $L_{\rm d} \lesssim 5 \times10^{-3} \, L_{\rm Edd}$ following the classification suggested by \cite{Ghisellini&etal_2011}.  Therefore, we take $L_{d} = 5 \times 10^{-3} L_{\rm Edd}$ as benchmark value for our sample. This corresponds to a photon energy density of $u_{\rm pl} \simeq 3 \times 10^{6} \, (M_{\bullet}/10^9 M_{\odot}) \, \rm erg \, cm^{-3}$ in the pair-plasma frame.

\subsubsection{The inner blob}
 Since this region is expected to be located close to the SMBH, we consider a compact size of a few $R_g$. Note that this scale can be associated with a minimum observed variability imposed by causality condition $t_{\rm var}^{\rm ob} \gtrsim r_g / \mathcal{D}_i \sim  \rm few \;  minutes$.\footnote{Rapid variability timescales of this order have been observed in some HBL, e.g., \cite{Aharonian&etal_2007ApJ...664L..71A,Albert&etal_2007ApJ...669..862A}.} The variability timescale in the VHE gamma-ray band is not established for the case of TBLs, except for 1ES 0229 +200 and 1ES 1218 +304; the former shows an indication of timescales of years and the last one on days. In that case, our model can explain the observation without violating the causality constraint even in the case of compact size of the order of $R'  \sim 10^{14} \, \rm cm \lesssim \mathcal{D}_i t_{\rm daily}$. 
 
\begin{figure*}
\begin{minipage}[b]{0.50\linewidth}
\centering
\includegraphics[width=0.9\linewidth]{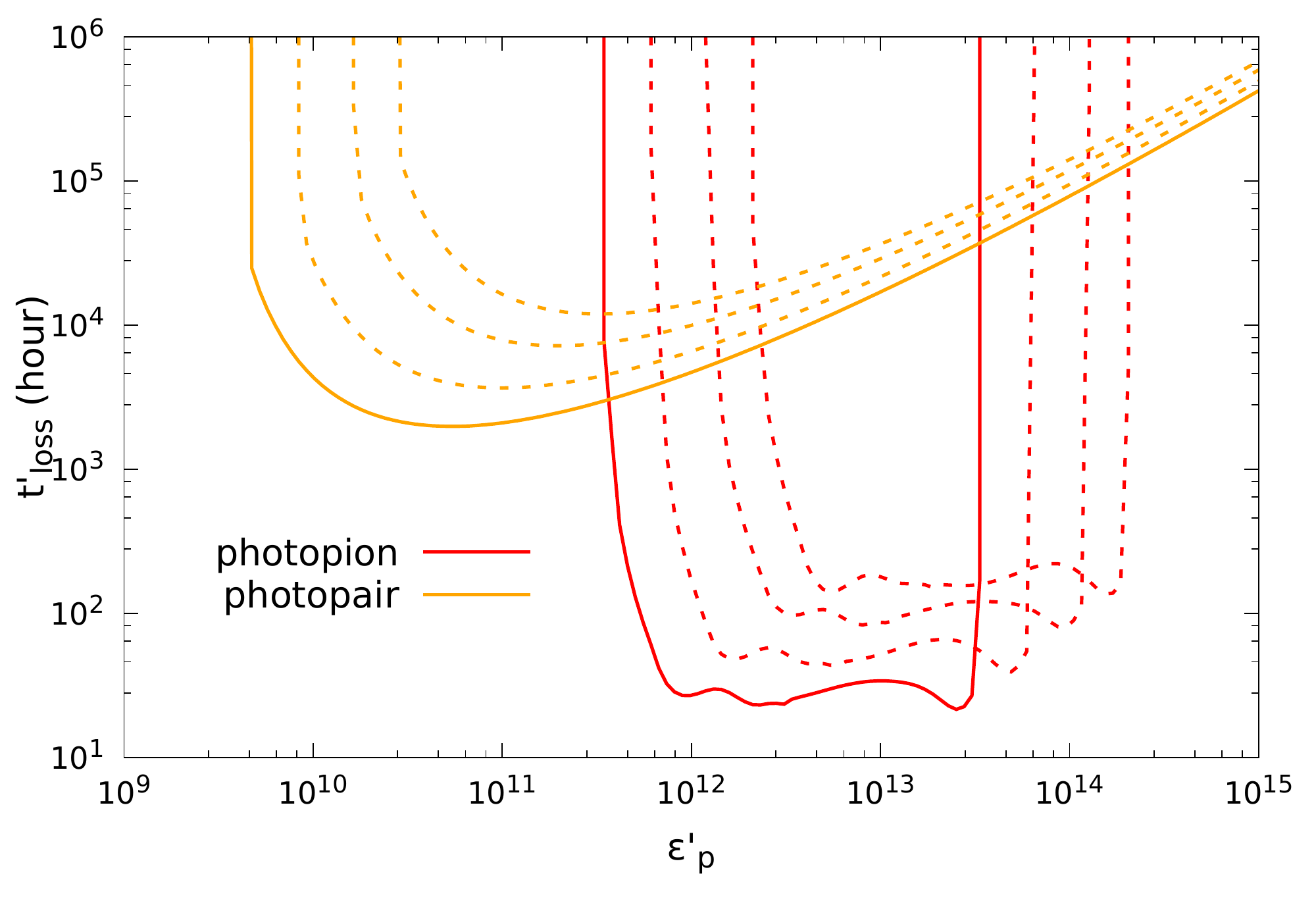} 
\end{minipage}\hfill 
\begin{minipage}[b]{0.50\linewidth}
\centering
\includegraphics[width=0.90\linewidth]{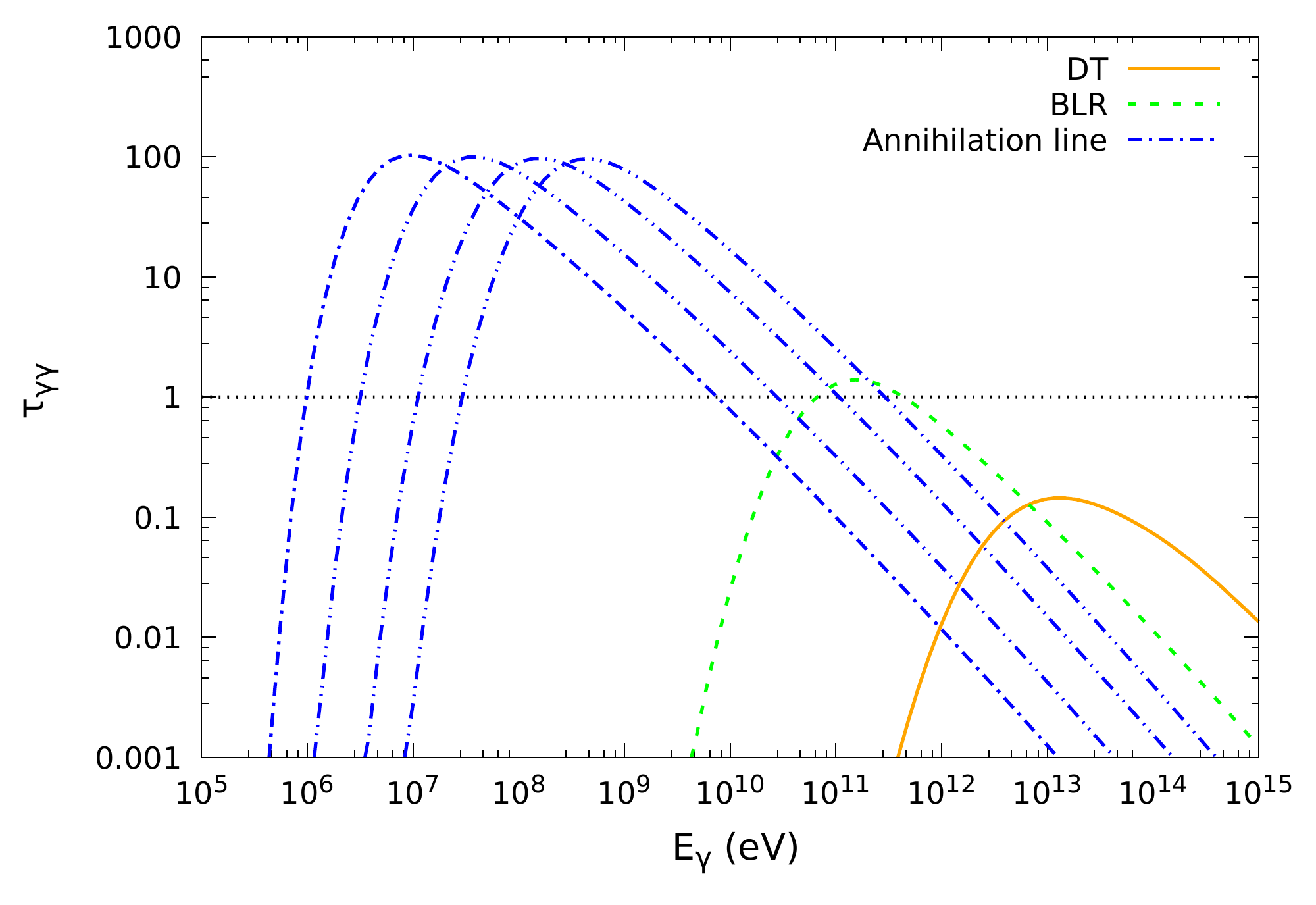} 
\end{minipage} 

\caption{ \textbf{Left:} Proton timescales in the comoving frame for photohadronic processes calculated in the inner blob. We consider external seed photons coming from the pair-plasma assuming a SMBH mass of $M_\bullet = 10^9 M_\odot$, an annihilation line luminosity $L_{\rm keV} =  5\times10^{-3}L_{\rm Edd}$, plasma velocity of $\beta_{\rm pl}=0.3$ and the boost Lorentz factor of the blob $\Gamma_{i}=1.5$ (solid line) and $\Gamma_{i}=3, 6, 10$ (dotted line from left to right). \textbf{Right:} Optical depth for $\gamma\gamma$ absorption produced in the inner blob considering different attenuation sources. In the case of the pair-plasma we use the same values of the process timescales with $\Gamma_i = 1.5 - 10$ from left to right.}
\label{fig_optdepth}
\end{figure*}

The hadronic contribution from this blob is calculated assuming a proton distribution with a hard spectral index of $\alpha_p=1.8$, being this value in the range of predicted by shock acceleration \citep{Stecker2007ApJ...667L..29S}. Also, we assume the minimum and maximum proton energies in the comoving frame as $\varepsilon'_{p,\rm min} = 1 \, \rm GeV$ and $\varepsilon'_{p,\rm max} = 100 \, \rm TeV$, respectively. Here, we do not assume a larger value of proton energy because the photons coming from the pair-plasma only interact via photopion process with protons up to that value,  although we use a Lorentz factor of $\Gamma_i = 10 $ (see Figure \ref{fig_optdepth}a). Also, this Figure shows that secondary electrons produced by the photopair process can be neglected compared with the photopion process contribution because the energy loss timescale is approximately two orders of magnitude greater.
As observed in Figure \ref{fig_optdepth}a), the resulting spectrum will be a narrow shape that shifts the peak to higher energies and decreases the timescale depending on $\Gamma_i$. As mentioned, we chose a hard proton distribution, then $\Gamma_i$ cannot be higher than 3 because higher values imply a $\gamma$-rays peak beyond 10 TeV. This statement is easy to see from the approximation $\varepsilon_\gamma = 0.1 \varepsilon_p$. As the efficiency is approximately constant in a specific proton energy range, the peak will be near that range's maximum proton energy value. For example, for a high value of $\Gamma_i$ we have $E^{\rm ob}_\gamma \simeq 100 \, {\rm TeV} \, (\varepsilon'_p/100 \, {\rm TeV}) (\Gamma_i/10)$, and  for a low value we have $E^{\rm ob}_\gamma \simeq 10 \, {\rm TeV} \, (\varepsilon'_p/30 \, {\rm TeV}) (\Gamma_i/3)$. From Table \ref{tab_Obs_parameters}, we note that 1ES 0229+200 is the only TBL whose expected peak is higher than 10 TeV, whereas the rest could be at lower energies, implying that $\Gamma_i$ could be less than 3. Therefore, we expect the Lorentz factor lies in the range of $1.5\leq \Gamma_i\leq 3$ for TBL sources. It is worth noting that the photopion process involves neutrino emission via charged pion decay products. 
Although this topic is beyond the scope of the current manuscript, a  simple estimation could be computed by using the  gamma-ray luminosity ($L_{\nu} \simeq \frac13 L_{\gamma}$) and spectral break ($\varepsilon_\nu \simeq 0.5 \varepsilon_\gamma$) \citep[ e.g.,][]{2014PhRvD..90b3007M, Fraija_2018MNRAS.481.4461F}. For example, for the case of 1ES 0229 +200 and using the values of table \ref{tab_EHBL_list}, we obtain $L_\nu \approx 5 \times 10^{44} \, \rm erg \, s^{-1}$ and $\varepsilon_\nu \approx 6 \, \rm TeV$. In a forthcoming paper, we will present a detailed study of neutrino emission of all TBLs. Here, we only present the neutrino flux of 1ES 0229 +200 as representative case (see the brown line of the left-hand top panel in Figure \ref{fig_SED_result}).
Furthermore, gamma-rays emitted in this region will be attenuated by the radiation of the annihilation line, BLR, outer-blob, and DT before being observed.  Figure \ref{fig_optdepth}b shows that the whole emission in the range of 1 MeV-10 GeV for $1.5\leq \Gamma_i\leq 3$ is strongly attenuated by the annihilation line emission; therefore, the cascade emission of secondary pairs would be suppressed. Additionally, it is essential to mention that although the BLR emission can be reflected in the inner blob spectrum, it has little attenuation effect, mainly at 100 GeV for $M_{\bullet} = 10^9 M_{\odot} $.  This effect would be smaller by a factor $\sim 3$ for $M_{\bullet} = 10^8 M_{\odot} $ compared to one with $M_{\bullet} = 10^9 M_{\odot} $ due to the scales of the optical depth $\tau_{\gamma\gamma} \propto r_{\rm BLR}$ and the radius $r_{\rm BLR} \propto L_d^{0.5} \propto L_{\rm Edd}^{0.5} \propto M_{\bullet}^{0.5}$. Therefore, only in the case of 1ES 0229+200 the effect is noticeable below TeV energies and then gamma-rays above $1 \, \rm TeV$ emerge without significant attenuation. It is worth noting that the DT does not play an attenuation effect at the highest energies.

The strength of the magnetic field in the inner blob could be estimated assuming that the magnetic luminosity is conserved along the jet, which is supported by observations \citep[e.g., see][]{2009MNRAS.400...26O}. Furthermore, \cite{1981ApJ...243..700K} suggested that a magnetic field varying as $B\propto r^{-1}$ can explain the electromagnetic emission of most compact VLBI jets. Here, we use the following expression
\begin{equation}\label{eq_magnetic_field_jet}
    B_i \simeq B_{\rm o} \left( \frac{r_i}{r_{\rm o}} \right)^{-1} \approx 0.1 \, {\rm G} \left( \frac{r_i}{0.1 \rm \, pc} \right)^{-1} \, ,
\end{equation}
where the parameter values of the outer blob will be discussed in next sections.  Table \ref{tab_model_parameters} shows that our model demands values of $B_{\rm o}\approx (0.1-0.3 ) \rm \, G$  and $r_{\rm o}\approx 10^{17} \rm cm$, for a daily variability timescale. Then using these values and a distance of the inner blob of $r_i \approx 10^{14} \rm \, cm$, the magnetic field in the inner blob becomes $B_i \sim 10^{2} \rm \, G$.  Accelerated protons confined by an intense magnetic field in this blob cool down by synchrotron process.  Protons accelerated at ultra-high energies (a few $\sim {\rm EeV}$) might radiate photons with energies at a few of $\sim {\rm GeV}$, which are  strongly attenuated by the radiation of the pair-plasma (see Figure \ref{fig_optdepth}). We do not consider the treatment of the proton synchrotron scenario because its effect is small compared with the radiation of primary electrons. On the other hand, the electron distribution in the inner blob is assumed to be described by a power-law function with the same spectral index as the proton one, $\alpha_e = \alpha_p$ \citep{Abdo_2011ApJ...736..131A}. Therefore, assuming the charge neutrality condition,  the electron normalization constant $K_e$ can be estimated once we determine the proton content. Electrons are cooling mainly by adiabatic expansion and synchrotron losses, and only the second mechanism produces a steeper distribution such that $\alpha_{e,2} = \alpha_{e,1} +1 \approx 3 $.  The condition of equality between adiabatic ($t_{\rm ad} = 2R'/c$) and synchrotron losses, ($t_{\rm syn} = 6\pi m_e c / (\sigma_{\rm T} \, B_i'^2 \, \gamma_e'$) could estimate the energy break of the electron population as \citep{2015MNRAS.448..910C}
\begin{equation}
    \gamma'_{e, \rm br} \approx 7.5  \, \left( \frac{B'_i}{100 \, \rm G} \right)^{-2} \left( \frac{R_i'}{10^{14} \rm \, cm} \right)^{-1} \,.
\end{equation}
Note that there is a strong dependency with the value of $B_i$, and if this value increases, then the energy break of the electron population shifts to lower energies and viceversa. For instance, if the magnetic field has a value of hundreds of Gauss and electrons are confined inside small regions, the energy break must appear only at relativistic energies avoiding the ultra-relativistic regime $\gamma_{e, \rm br}' \gg 10$. Therefore, a strong magnetic field implies a steeper electron distribution, suppressing the synchrotron radiation. The observed energy peak of the electron synchrotron in this blob is determined by the value of the energy break, which is given by the above relation. Then, we obtain the expression
\begin{equation}  
    E_{e, \rm syn, pk}^{\rm ob} \approx 3 \times 10^{-4} \, {\rm eV} \, \left( \frac{\mathcal{D}_i}{5} \right) \left( \frac{B_i'}{100 \, \rm G} \right)^{-3} \left( \frac{R_i'}{10^{14} \rm \, cm} \right)^{-2}\,.
\end{equation}
We note that the synchrotron energy peak is very dependent on the strength of the magnetic field, with $B_i\sim 100 \rm \, G$, and the peak appears in the radio band.  Therefore, the data reported for some TBLs in the radio band can help us to constrain the value of the magnetic field via synchrotron radiation in this region
\begin{align}
 B'_i
 &\sim 
 \left( \frac{6\pi \nu L_{\nu}^{\rm ob}}{\sigma_{\rm T} c  K_e'} \right)^{\frac{2}{1+\alpha_{e,1}}} \left( \frac{m_e \nu}{e}\right)^{\frac{-3+\alpha_{e,1}}{1+\alpha_{e,1}}} {\mathcal{D}_i}^{\frac{-5-\alpha_{e,1}}{1+\alpha_{e,1}}}
\end{align}
For example, considering the case of 1ES 0229+200 with $\alpha_{e,1} = 2$ and $K_e'$ (determined by the neutrality condition), we obtain
\begin{equation}
    B'_i \approx 45 {\, \rm G} \, \bigg(\frac{\mathcal{D}_i}{6}\bigg)^{-7/3}  \bigg(\frac{\nu_{\rm syn}^{\rm ob}}{10 \, \rm GHz}\bigg)^{-1/3}  \bigg(\frac{\nu L_{\nu, \rm syn }^{\rm ob}}{10^{41} \, \rm erg \, s^{-1}}\bigg)^{2/3}  \bigg(\frac{K_e'}{10^{49}}\bigg)^{-2/3}\,
\end{equation}
which agrees with the suggested one by the equation (\ref{eq_magnetic_field_jet}).

\subsubsection{The outer blob}

We assume this blob is moving faster than the inner blob with $\Gamma_{\rm o} = 5$, and the blob's size is associated with a variability timescale of one day, corresponding to $R_{\rm o}'' \lesssim 2.6 \times 10^{16} \, {\rm cm} \, (t_{\rm var}/{\rm day})(\mathcal{D}_{\rm o}/10)$. If the blob's radius corresponds to the jet cross-section and the jet half-opening angle is $\theta_j \simeq 1/(\Gamma_{\rm o})$, then the dissipation distance could be estimated by the expression
\begin{equation}
    r_{\rm o} \simeq 2 \Gamma_{\rm o}^2 c t_{\rm var} \approx 1.3 \times 10^{17} \, {\rm cm} \, \left(\frac{\mathcal{D}_{\rm o}}{10}\right)^2 \, \left(\frac{t_{\rm var}}{{1 \, \rm day}}\right)\,.
\end{equation}
Moreover, using the equations (\ref{eq_rBL}) and (\ref{eq_rDT}) the location of the BLR and the DT is $r_{\rm BLR} \approx 7 (2)\times10^{16} {\rm cm}$ and $r_{\rm DT} \approx 1.7 (0.6) \times 10^{18} \, {\rm cm}$ for $M_{\bullet} = 10^9 (10^8) \, M_{\odot}$ and $L_d = 5\times 10^{-3} L_{\rm Edd}$, respectively. Note that the dissipation region is located outside the BLR, but still immersed in the DT radiation field $r_{\rm BLR} \leq r_{\rm o} \leq r_{\rm DT}$. Therefore, as the outer blob is beyond the BLR influence, its radiation can be discarded, and only the infrared DT photons will be considered in our calculation.

\paragraph{The broad line region.}\label{estimations_BLR} Discussion around the electromagnetic emission of BLR/DT  points out that some BL Lacs may have weaker BLR luminosity.  For example, EHSP BL Lacs like Mrk 421, Mrk 501 and H 2356-309 have BLR luminosities of $L_{\rm BLR}\approx 5 \times 10^{41} \rm \, erg \, s^{-1}$, $\approx 1.3 \times 10^{42} \rm \, erg \, s^{-1}$ \citep{Sbarrato_2012MNRAS.421.1764S} and $\sim 10^{41} \, \rm erg \, s^{-1}$ \citep{Fang_2014ApJ...795...57F}, respectively. All of these luminosity values observed around $L_{\rm BLR} \sim 10^{-5} L_{\rm Edd}$ were computed assuming SMBH masses \cite{Wagner_2008MNRAS.385..119W}. Furthermore, we note that the formation of pair-plasma demands a value of  $L_d \sim 10^{-3} L_{\rm Edd}$, which translates into a ratio of $\phi_{\rm BLR}= L_{\rm BLR}/L_d \sim 10^{-2}$. It is worth mentioning that for a standard accretion disk, it is well accepted a value of $\phi_{\rm BLR} = 0.1$. In this work, we adopt $\phi_{\rm BLR} = 0.05$ as the reprocessed radiation from the disc in the BLR. Then, the BLR luminosities must be $L_{\rm BLR} \approx 2\times 10^{\rm 43} (2\times 10^{\rm 42}) \rm \, erg \, s^{-1} \, $ for $M_{\bullet}=10^9 (10^8) M_{\odot} $. These values are in the range found for a EHSP BL Lacs with a possible high-energy neutrino association $L_{\rm BLR}<3 \times 10^{-4} \, L_{ \rm Edd}$ \citep{Giommi&etal_2020}. From equation (\ref{eq_rBL}), the BLR location is $r_{\rm BL} \approx 7 \times 10^{16} \, {\rm cm} \, ( M_{\bullet}/10^9 M_{\odot} )^{0.5}$ and the energy density is 
\begin{equation}
    u''_{\rm BL} \approx 9.5 \times 10^{-5} \, \frac{\rm erg }{\rm cm^{3}}\, \left( \frac{\mathcal{D}_o}{10}\right)^{-2} \, \left(\frac{\phi_{\rm BL}}{0.05}\right)\, ,
\end{equation} 
with line's shape spectrum peaking at $\varepsilon''_{\rm BL} \approx 1 \, (\mathcal{D}_o/10)^{-1} \, \rm eV$.

\paragraph{The dusty torus.}\label{estimations_DT} The DT is assumed with the same fraction of the disc luminosity as the BLR, i.e., $\phi_{\rm DT} = 0.05 $. Using equations (\ref{eq_rDT}) and (\ref{eq_DTenergy}), we can see that the DT radius is $r_{\rm DT} \approx 10^{18} \, {\rm cm} \, ( M_{\bullet}/10^9 M_{\odot} )^{0.5}$, and the energy density is 
\begin{equation}
    u''_{\rm DT} \approx 2.1 \times 10^{-2} \, \frac{\rm erg }{\rm cm^{3}}\, \left( \frac{\mathcal{D}_o}{10}\right)^{2} \, \left(\frac{\phi_{\rm DT}}{0.05}\right)\, ,
\end{equation}
and spectrum peaks at an energy of $\varepsilon''_{\rm DT} \approx 1 \, (\mathcal{D}_o/10) \, \rm eV$. Then, we expect EC will be mainly dominated by the DT contribution instead of the BLR.  

\paragraph{Synchrotron radiation.}\label{estimations_sycnhrotron}
The relativistic electrons in this region are responsible for the X-rays emission via synchrotron radiation.  To explain the strength of magnetic field  in the range of $0.1 - 1 \, \rm G$, we use a maximum value of the break electron Lorentz factor ($\gamma_{e,\rm b}^{\prime\prime}= 4\times10^5$) and  the observed peak energy with the average value  $E_{s, \rm pk}^{\rm ob} \sim  3 \, {\rm keV}$ (see Table \ref{tab_EHBL_list}). In this case, the strength of magnetic field is
\begin{equation}
 B_o'' \gtrsim 0.1 \, {\rm G} \, (1+z) \, \left( \frac{E_{\rm s}^{\rm ob}}{{3 \, \rm keV}} \right)  \left(\frac{\mathcal{D}_o}{10}\right)^{-1} \left(\frac{\gamma_{e,\rm b}^{\prime\prime}}{4\times10^5}\right)^{-2}\, .
\end{equation}
Furthermore, the energy density of synchrotron emission produced by this blob, assuming a variability timescale of one day is
\begin{align}
    u_s^{\prime\prime} 
    &= 
    L_s^{\rm ob}/ (4\pi \mathcal{D}_{\rm o}^4 {R_{\rm o}^{\prime\prime}}^2 c) 
    \\
    &\sim 
    10^{-2} \, {\rm \frac{erg}{cm^{3}}} \, \left( \frac{\mathcal{D}_o}{10} \right)^{-4} \left( \frac{R_{\rm o}^{\prime\prime} }{10^{16} \, \rm cm} \right)^{-2} \, \left( \frac{L_s^{\rm ob} }{10^{45} \, \rm erg \, s^{-1}} \right) \,.
\end{align}
For an electron distribution with power-law and index $p\sim 2$, the corresponding normalization constant at the peak is approximated by
\begin{align}
    K_e^{\prime\prime}
    &\sim 
    6\pi L_{\rm s, pk}^{\rm ob} / ( \sigma_{\rm T} c \, \mathcal{D}_o^4 {B_o''}^2 \, {\gamma_{e,b}^{\prime\prime}}^{3-p} ) 
    \\
    &\sim 8 \times 10^{51} \,  \left( \frac{\mathcal{D}_o}{10} \right)^{-4} \left( \frac{B_o^{\prime\prime}}{0.5 \, \rm G} \right)^{-2} \, 
    \left( \frac{\gamma_{e,b}^{\prime\prime} }{2 \times 10^{5}} \right)^{-1} \left( \frac{L_{\rm s, pk}^{\rm ob} }{10^{45} \, {\rm erg \, s^{-1} }} \right) \, .
\end{align}

\paragraph{High-energy emission.} All electrons with $\gamma_e^{\prime\prime}  \ll 5\times 10^5 ( \epsilon_{\rm DT}/ 0.1 {\rm eV} )^{-1} (\mathcal{D}_o/10)^{-1} $ scatter off in the Thompson regime with the DT radiation, avoiding suppression due to KN regime. For instance, if we choose $\gamma_{e,\rm br}^{\prime\prime}  = 2\times 10^5$, then the peak of EC is expected around $E_{ic}^{\rm ob} \sim 0.48 \, {\rm TeV} (\gamma_e^{\prime\prime  } / 2 \times 10^5)^2 (\mathcal{D}_o/10)^2 (\epsilon_{\rm DT}/0.1 \, {\rm eV})$; just below the gamma-rays peak produced by photopion in the inner blob. 

The electron distribution is assumed as a broken power-law function with free parameters: the spectral indexes $\alpha_{e,1}$ and $\alpha_{e,2}$, and the minimum ($\gamma_{e, \rm min}$), the break ($\gamma_{e, \rm b}$), and the maximum ($\gamma_{e, \rm max}$) electron Lorentz factor. Also, the magnetic field, $B_o$, and the size of the blob, $R_o$, are used to fit the data. 

Finally, using the parameters listed in Table \ref{tab_model_parameters} we model successfully the broadband SED of six known \textit{extreme} BL Lac 1ES 0229+200, RGB J0710+591, 1ES 0347-121, 1ES 1101-232, 1ES 1218-304, 1ES 0414+009. The result of our model is shown in Figure (\ref{fig_SED_result}). It is important to mention that in this work, we have not included the contribution observed in the optical band produced by the host galaxy \citep{Costamante2018}.

\section{Discussion and conclusions}

The \textit{extreme} BL Lac behavior presents challenges for a pure leptonic or hadronic model with one-zone emission. These objects demand atypical parameters for explaining the hard spectrum shape, especially in a VHE gamma-ray band. In this work, we proposed a lepto-hadronic model with two-zone emission to explain the multi-wavelength observations for the so-called extreme blazars both in synchrotron and TeV component.  In particular cases,  we described the six best-known \textit{extreme} BL Lacs using our proposed model, and we found a feasible explanation that relaxes the atypical values required by the one-zone models.

\begin{figure*}
\begin{minipage}[b]{0.50\linewidth}
\centering
\includegraphics[width=1.01\linewidth,height=.23\textheight]{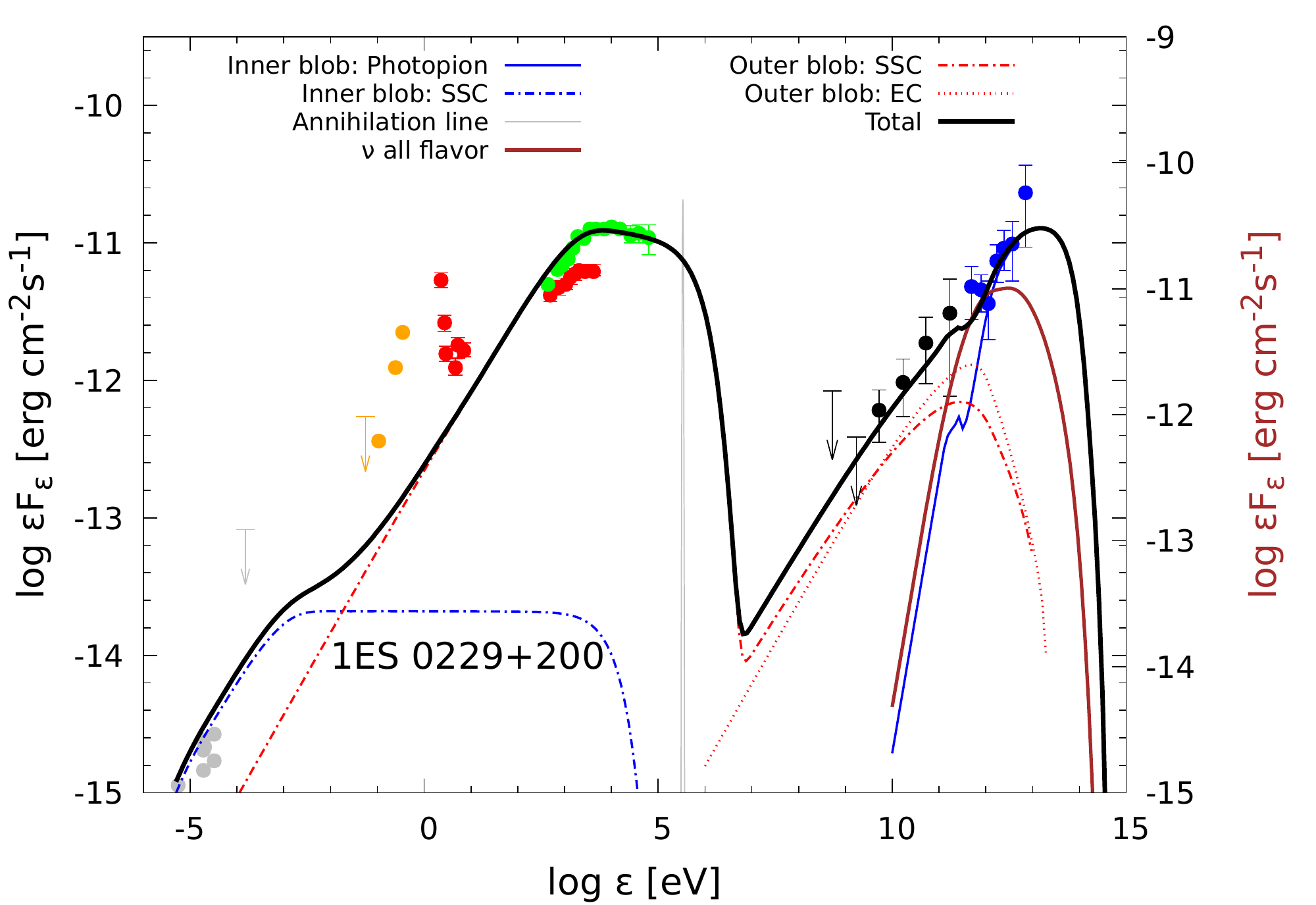}
\end{minipage}\hfill 
\begin{minipage}[b]{0.50\linewidth}
\centering
\includegraphics[width=0.90\linewidth]{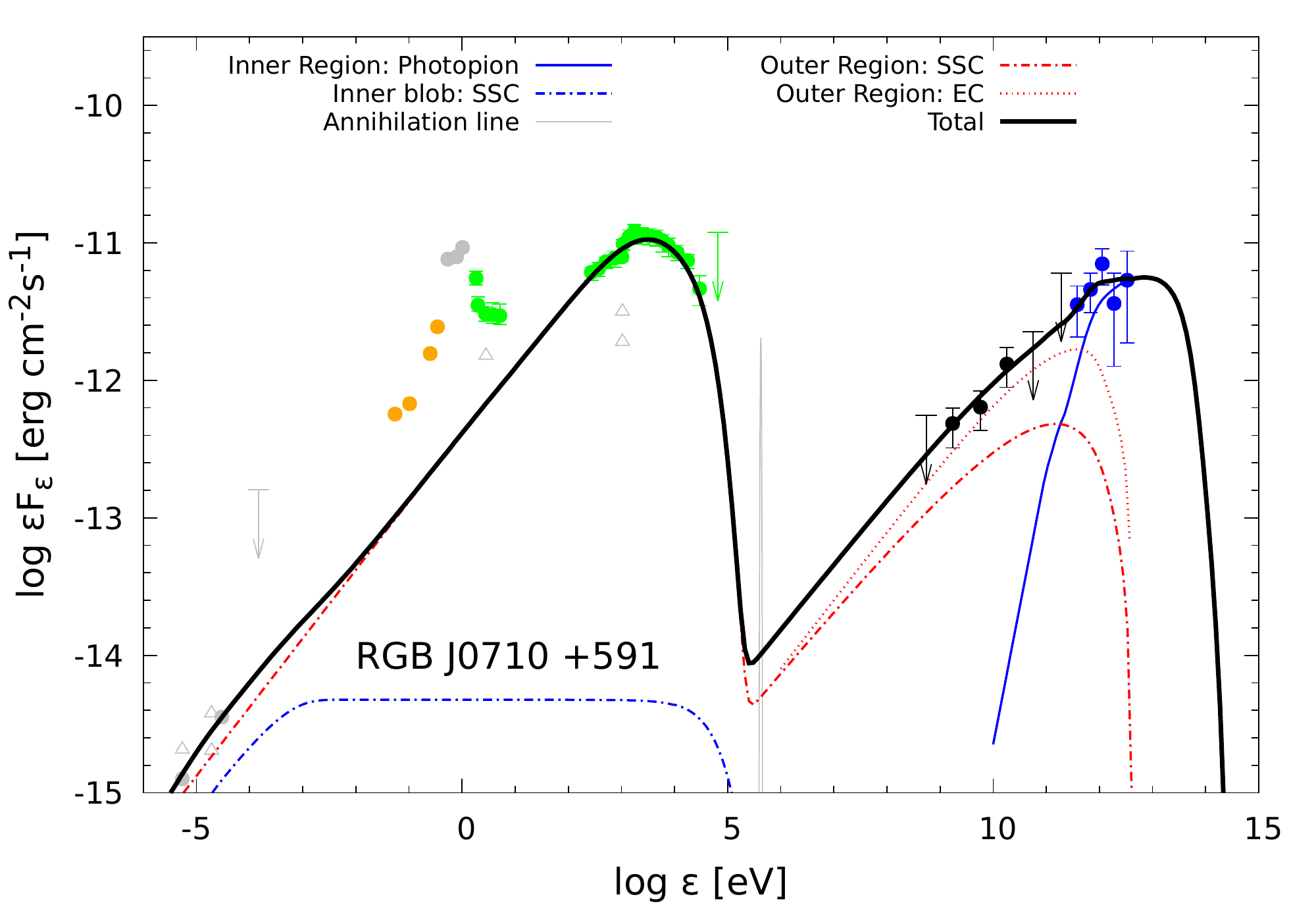} 
\end{minipage} 
\begin{minipage}[b]{0.50\linewidth}
\centering\includegraphics[width=0.90\linewidth]{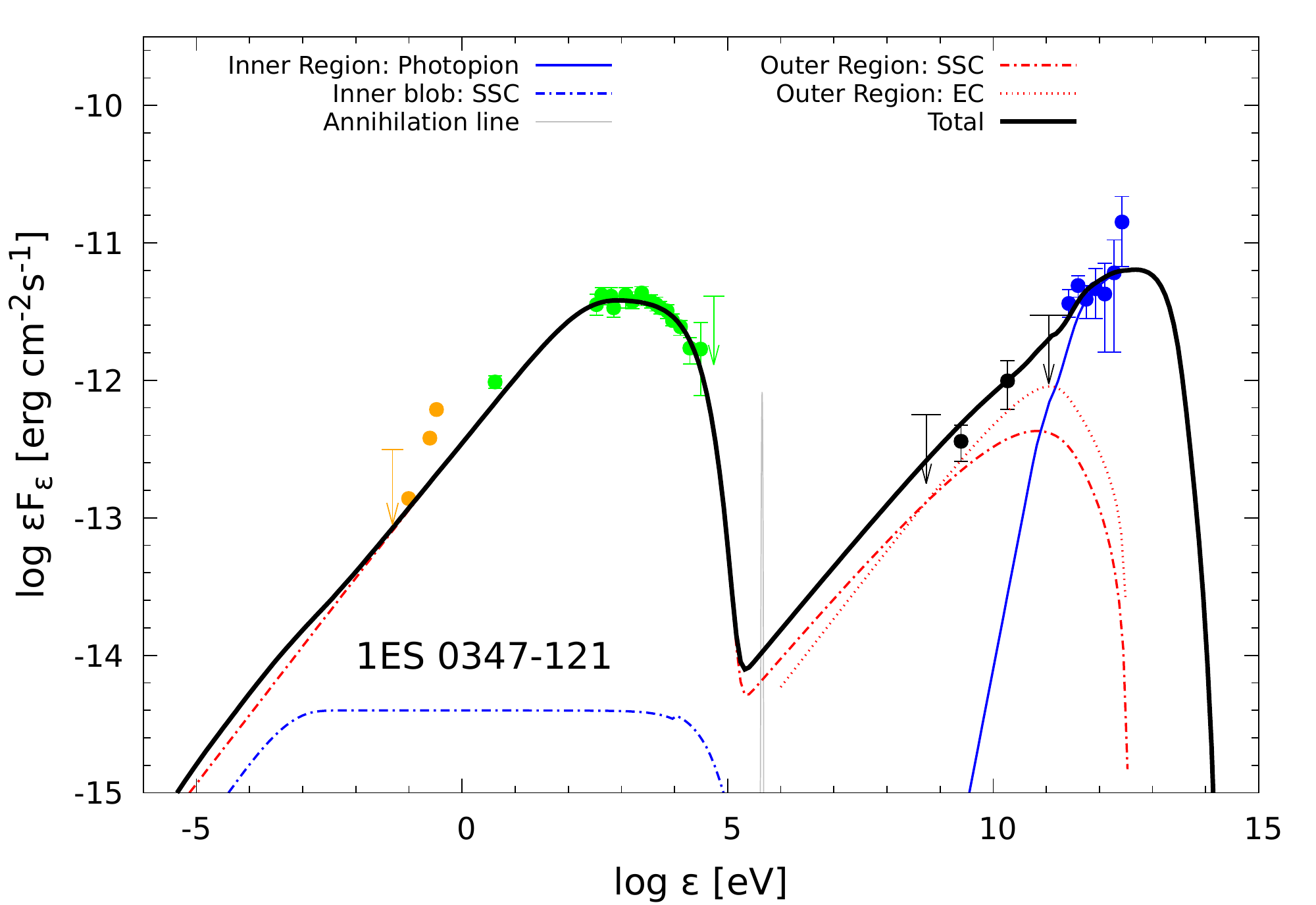}
\end{minipage}\hfill 
\begin{minipage}[b]{0.50\linewidth}
\centering\includegraphics[width=0.90\linewidth]{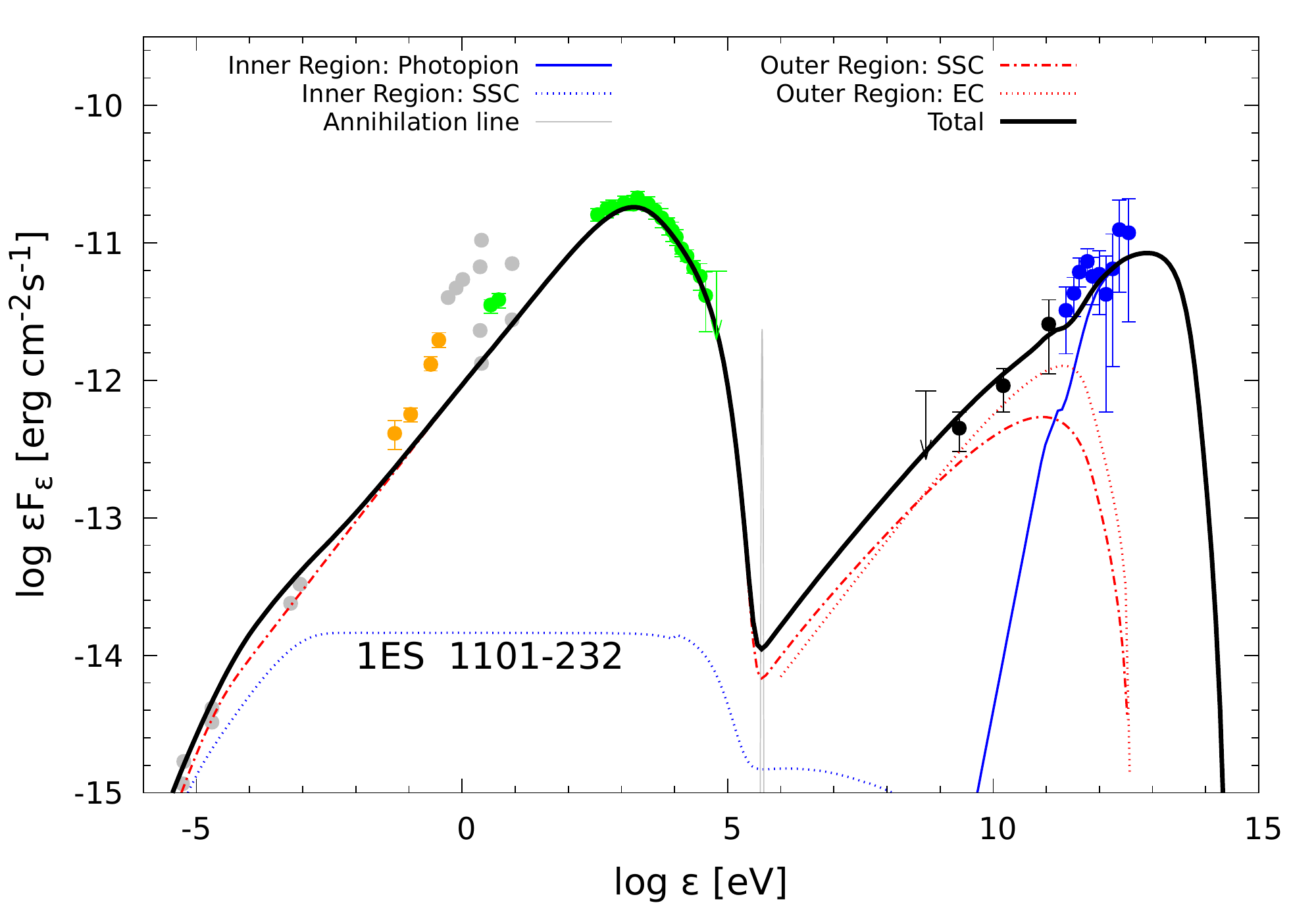}
\end{minipage} 
\begin{minipage}[b]{0.50\linewidth}

\centering\includegraphics[width=0.90\linewidth]{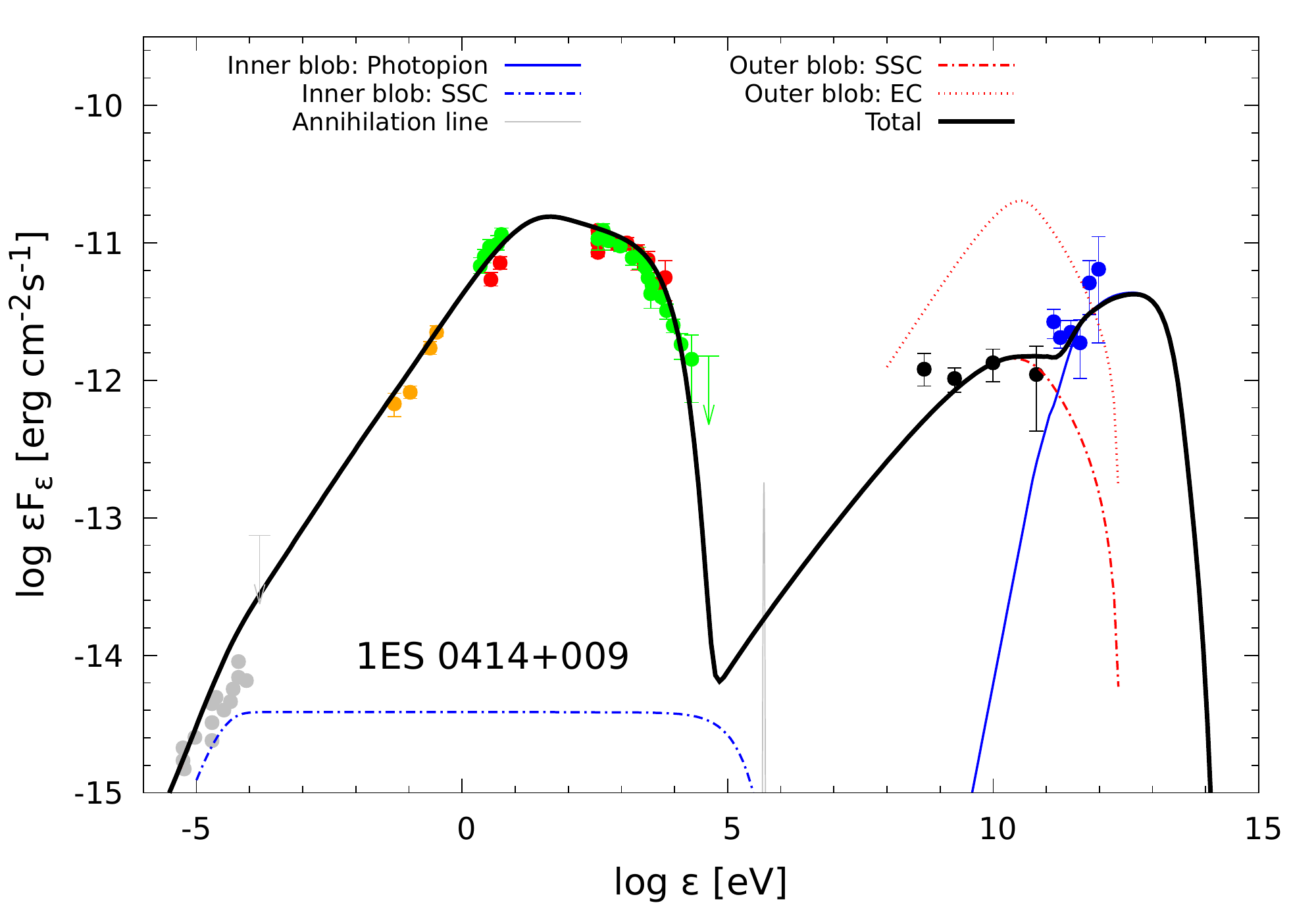}
\end{minipage}\hfill 
\begin{minipage}[b]{0.50\linewidth}
\centering\includegraphics[width=0.90\linewidth]{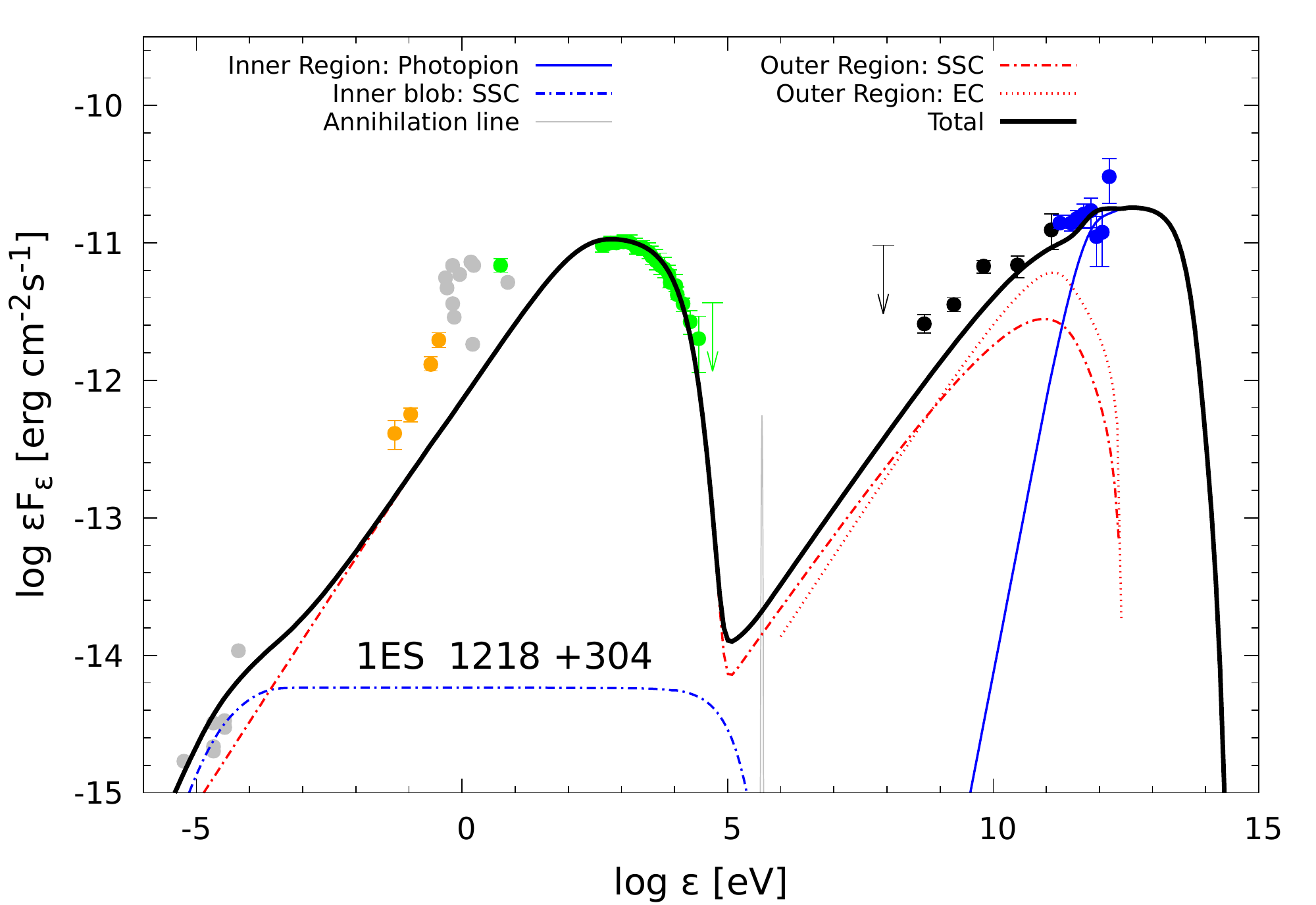}
\end{minipage}\hfill 
\caption{The broadband SED of the six best-known \textit{extreme} BL Lacs with the two-zone lepto-hadronic model proposed in this work. The SSC model and the EC model from the outer blob  with seed photons from the DT  are shown in dotted-dashed and dotted red lines. Photo-pion decay products above at TeV energies from the inner blob is shown in solid blue line. The sum of all contributions is shown with solid black lines. The brown line of the left-hand top panel (i.e., 1ES0229 + 200) corresponds to the neutrino flux.  Data are from \protect\cite{Costamante2018} and see references therein.}
\label{fig_SED_result}
\end{figure*}

We showed that the annihilation line radiation produced in a mildly relativistic outflow pair-plasma and the DT play essential roles in forming the hard spectrum observed in the VHE gamma-ray bands.   In our model, this gamma-ray bump is produced mainly in two regions: i) the inner blob, which is responsible for the highest energy contribution above TeV-energies via photo-hadronic interactions with the annihilation line photons from the pair-plasma, and ii) the outer blob, where the very-high-energy gamma rays are created by EC with seed photons from the DT, and by the SSC mechanism.  In all cases the inner blob is a compact region of the order of $R_i \sim 10^{14} \, \rm cm$ which moves with a low boost factor, i.e, $\Gamma_i = 1.5-3 \, (\mathcal{D}_i = 2.6-5.8)$ and with a proton luminosity in the sub-Eddington regime which is a feasible value for BL Lacs. The outer blob can explain the whole X-rays emission and sub-TeV flux with a SSC model without demanding an equipartition value far from one. Our result shows that $U_B/U_e \gtrsim 0.1$.  Considering the effects of external radiation from the DT, which is the dominant contribution by a factor of $\sim 2$, a harder spectrum can be obtained in the GeV band if otherwise only synchrotron photons are considered. Only the EHSP BL Lac 1ES 0414 +009, which seems most likely a HSP BL Lac (because its peak is $\approx$ 0.4 keV), is well explained without the DT contribution (see Figure \ref{fig_SED_result}). This could mean that the outer blob is located outside from the DT radiation influence. Therefore, the size would be bigger than those which outer blob are inside the DT. Our result is consistent with this idea because of 1ES 0414 +009 has a outer blob size bigger than the size of the BLR ($R_{\rm BLR} \sim 10^{16} \, \rm cm$) as can see in Table (\ref{tab_model_parameters}).

Another significant result is that the electron-synchrotron process originating from the inner blob could explain the radio observations, which for the case of 1ES 0229 +200, can hardly be interpreted by any radiative mechanism in the outer blob.

It is crucial that the outer blob still requires a high value of the break electron Lorentz factor $\gamma_{\rm e,b}^{\prime\prime}\approx10^5$ to explain the synchrotron peak and part of the gamma-ray spectrum. As many authors have pointed out, this result implies that extreme blazars must be very efficient particle accelerators.  Furthermore,  our model does not require a high value of minimum electron Lorentz factor in the outer blob. The maximum value required in our model is $\approx$ 50, in contrast with the typical values around $10^3$, as demanded in SSC models. Finally, it is essential to notice that if either the pair-plasma or the inner blob are not present, the EHSP behavior is conserved although the TBL is turned off. This result agrees with the existence of EHSP without exhibiting TBL behavior or vice-versa. 

\section*{Data availability}
The data used in this article are available in TeVCat Website, at http://tevcat2.uchicago.edu/

\section*{Acknowledgements}
We are grateful to Antonio Marinelli for useful discussions.  NF and EB acknowledges  financial support  from UNAM-DGAPA-PAPIIT  through  grants  IN106521 and IN113320.

\bsp	
\label{lastpage}

\bibliographystyle{mnras}
\bibliography{references.bib}

\clearpage

\begin{table*} 
\caption{
Observational parameters for the six best-known \textit{extreme}BL Lacs listed in \citet{Costamante2018} and used in this work.
Column(1): Object name.
Column (2): The log SMBH in solar units from \citet{Woo_2005ApJ...631..762W} (for RGB J0710 +591) and from \citet{Wagner_2008MNRAS.385..119W} (for the remaining objects). 
Column (3): Luminosity distances derived in this work. 
Column (4): Redshifts provided in \citet{Costamante2018}. 
Column (5): Synchrotron energy peaks reported in \citet{Costamante2018}. 
Column (6): Synchrotron luminosities at different energy peaks calculated with the luminosity distance and the synchrotron flux provided in \citet{Costamante2018}. 
Column (7): Lower limits of the energy peaks at VHEs \citet{Costamante2018}. 
Column (8): Lower limits of VHE luminosities at the energy peaks.
Column (9): The variability timescales at VHEs.
}
\label{tab_Obs_parameters}
\begin{tabular}{l|c|c|c|c|c|c|c|c}
\hline
 Object name & $\log(M_{\bullet})$ &  $d_L$  & $z$ & $E_{\rm syn}^{pk}$ & $\nu L_{\nu,\rm syn}^{pk}$ & $E_{\rm VHE}^{pk}$ & $\nu F_{\nu, \rm VHE}^{pk}$
 \\
  & $(M_\odot)$ & $(\rm Gpc)$ &  & (keV) & $(\rm erg \, cm^{-2} \, s^{-1})$ & (TeV) & $(\rm erg \, s^{-1})$ & Variability at VHEs 
\\
$[1]$ & [2] & [3] & [4] & [5] & [6] & [7] & [8] & [9]
\\
\hline
\\
\textbf{1ES 0229 +200 }  & $9.16\pm 0.11$   &   0.643 & 0.140 & $9.1\pm 0.7$ & $6.48 \times 10^{44}$ & $ > 12 $ & $ > 9.8 \times 10^{44}$ & years {$^a$}
\\
\textbf{RGB J0710 +591 } & $8.25\pm 0.22$   &   0.569 & 0.125 & $3.5\pm 0.2$ & $4.3 \times 10^{44}$ & $> 4 $ & $> 2.3 \times 10^{45}$ & unknown
\\
\textbf{1ES 0347-121}    & $8.02\pm 0.11$   &   0.899 & 0.188 & $1.4\pm 0.6$ & $4.03 \times 10^{44}$ & $> 3 $ & $> 7.7 \times 10^{44}$ & unknown
\\
\textbf{1ES 1101-232}    & $8.4^c$              &   0.879 & 0.186 & $1.4\pm 0.2$ & $1.83 \times 10^{45}$ & $> 4 $ & $> 9.2 \times 10^{44}$ & unknown
\\
\textbf{1ES 0414 +009}    & $8.4^c$              &   1.435 & 0.287 & $0.3\pm 0.2$ & $2.78 \times 10^{45}$ & $> 2 $ & $> 7.3 \times 10^{44}$ & unknown
\\
\textbf{1ES 1218 +304}    & $8.04 \pm 0.24$              &   0.858 & 0.182 & $1.3 \pm 0.3$ & $7.9 \times 10^{44}$ & $> 2 $ & $> 1.8 \times 10^{45}$ & days $^b$
\\
\hline
\end{tabular}\label{tab_EHBL_list}
   \\Reference: $^a $ \cite{Aliu&etal_2014ApJ...782...13A}; $^b$  {\cite{Acciari&etal_2010ApJ...709L.163A}};
   \\ \textbf{ Note:} $^c$ This value is assumed considering the average one of the rest objects is $8.36\pm0.17$.
\end{table*}

%
%
%
%


%
%
%
%

\begin{table*}
\caption{Parameters used to describe the broadband SED of the six best-known \textit{extreme} BL Lacs.}
\begin{tabular}{l|c|c|c|c|c|c|c|c}
\hline
  &  \textbf{1ES 0229 +200 } & \textbf{ RGB  J0710 +591 } & \textbf{ 1ES 0347-121}  & \textbf{1ES 1101-232}  & \textbf{1ES 0414 +009}  & \textbf{1ES 1218 +304}
\\
\hline
\textbf{ Inner Blob }
\\
\textbf{$\Gamma_i (\mathcal{D}_i)$}      & 3(5.8)    & 2.5(4.8) & 1.5(2.6)   &   2(3.7) & 1.5(2.6) & 2(3.7)
\\
\textbf{$R_b$}[$10^{14} \, \rm cm$]  & 2.3  & 0.76 & 0.76 & 0.76 & 0.7  & 0.76      
\\
\textbf{$K_p$} [$\rm eV^{-1} \, cm^{-3}$]        
    & $3.9 \times 10^{-4}$  
    & $2.5 \times 10^{-3}$   
    & $2.2 \times 10^{-2}$ 
    & $1.7 \times 10^{-2}$ 
    & $2.7 \times 10^{-2}$ 
    & $1.0 \times 10^{-2}$
\\
\textbf{$L_p \rm \; [erg \, s^{-1}]$} 
    & $8.9 \times 10^{45}$ 
    & $1.2 \times 10^{45}$ 
    & $3.4 \times 10^{45}$ 
    & $1.3 \times 10^{46}$ 
    & $3.7 \times 10^{45}$ 
    & $5.1 \times 10^{45}$
\\
\textbf{$L_e \rm \; [erg \, s^{-1}]$} 
    & $7.1 \times 10^{41}$ 
    & $1.4 \times 10^{41}$ 
    & $2.5 \times 10^{41}$ 
    & $9.1 \times 10^{41}$ 
    & $1.21 \times 10^{41}$ 
    & $2.3 \times 10^{41}$
\\
\textbf{$L_B \rm \; [erg \, s^{-1}]$} 
    & $4.0 \times 10^{43}$ 
    & $4.1 \times 10^{42}$ 
    & $1.1 \times 10^{42}$ 
    & $3.4 \times 10^{42}$ 
    & $2.3 \times 10^{43}$ 
    & $1.4 \times 10^{43}$
\\
\textbf{$\gamma_{e,\rm b}$}   
    & $17$      & $11$  
    & $13$      & $13$ 
    & $1$      & $3.3$

\\
\textbf{$B \; \rm [G]$ }   & 50   & 110  &  $100$   &   100  & 500  & 200
\\
\textbf{$U_B/(U_p+U_e)$}& 0

.004& 0.003 & $3.2 \times 10^{-4}$ & $2.3 \times 10^{-4}$ & 0.006 & 0.003
\\
\\
\textbf{ Outer Blob }
\\

\textbf{$B \rm \; [G]$}     & 0.18 & 0.16 & 0.18 & 0.3 & 0.1 & 0.15

\\
\textbf{$R_b$ [$10^{16} \, \rm cm$]}  & 1.3 & 1.9 & 1.9 & 3.4 & 23 & 2.6
\\
\textbf{$\gamma_{\rm min}$}   & $1$  & $1$  & $1$ & $30$ & $50$ & $1$
\\
\textbf{$\gamma_{\rm b}$}   
    & $3\times 10^{5}$  
    & $3\times 10^{5}$  
    & $1\times 10^{5}$ 
    & $2\times 10^{5}$ 
    & $4\times 10^{4}$ 
    & $1\times 10^{5}$
\\
\textbf{$\gamma_{\rm max}$} 
    & $5\times 10^{6}$    
    & $1\times 10^{6}$  
    & $9\times 10^{5}$ 
    & $1\times 10^{6}$ 
    & $7\times 10^{5}$ 
    & $7\times 10^{5}$
\\
\textbf{$\alpha_{\rm e,1}$} & 1.8 & 2   & 2 & 2   & 1.8 & 1.8
\\
\textbf{$\alpha_{\rm e,2}$} & 3.1 & 3.2 & 3 & 3.9 & 3.2 & 3
\\
\textbf{$L_B \rm \; [erg \, s^{-1}]$}       
    & $1.27 \times 10^{43}$ 
    & $2.26 \times 10^{43}$ 
    & $2.86 \times 10^{43}$ 
    & $2.39 \times 10^{44}$ 
    & $1.2 \times 10^{45}$ 
    & $3.5 \times 10^{43}$ 
\\
\textbf{$L_e \rm \; [erg \, s^{-1}]$} 
    & $6.2 \times 10^{43}$ 
    & $9.67 \times 10^{43}$ 
    & $1.58 \times 10^{44}$ 
    & $8.17 \times 10^{43}$ 
    & $4.6 \times 10^{44}$ 
    & $2 \times 10^{43}$ 
\\
\textbf{$U_B/U_e$}  & 0.2 & 0.23 & 0.18 & 2.9  & 2.7 & 0.18
\\
\\
\hline
\end{tabular} \label{tab_model_parameters}
\end{table*}

\end{document}